\pdfoutput=1
\documentclass[iop]{emulateapj}

\usepackage{epstopdf}
\usepackage{lineno}
\usepackage{xspace}
\usepackage{graphicx}
\usepackage{amssymb}
\usepackage{amsmath}
\usepackage{natbib}
\usepackage{float}
\usepackage{hyperref}

\shorttitle{Jet production efficiency in young radio galaxies}
\shortauthors{W\'ojtowicz et al.}

\begin{document}

\title{On the Jet Production Efficiency in a Sample of the Youngest Radio Galaxies}

\author{A. W\'ojtowicz \altaffilmark{1}, \L .~Stawarz \altaffilmark{1}, C.~C.~Cheung \altaffilmark{2}, L.~Ostorero \altaffilmark{3}, E.~Kosmaczewski \altaffilmark{1} \& A.~Siemiginowska \altaffilmark{4}}

\altaffiltext{1}{Astronomical Observatory of the Jagiellonian University, ul. Orla 171, 30$-$244 Krak\'ow, Poland}
\email{email: {\tt: awojtowicz@oa.uj.edu.pl}}
\altaffiltext{2}{Naval Research Laboratory, Washington, DC 20375, USA}
\altaffiltext{3}{Dipartimento di Fisica -- Universit\'a degli Studi di Torino and Istituto Nazionale di Fisica Nucleare (INFN), Via P. Giuria 1, I-10125 Torino, Italy}
\altaffiltext{4}{Harvard Smithsonian Center for Astrophysics, 60 Garden Street, Cambridge, MA 02138, USA}

\begin{abstract}
Here we discuss the jet production efficiency in a sample of 17 young radio galaxies with measured redshifts, kinematic ages, and nuclear X-ray fluxes, for which the observed luminosities of compact jets/lobes and accretion disks correspond to the same episode of the AGN activity. For the targets, we analyze the available optical data, estimating the bolometric luminosities of the accretion disks $L_{\rm bol}$, and the black hole masses; we also derive the minimum jet kinetic luminosities, $P_{\rm j}$. With such, we investigate the distribution of our sample in the three-dimensional space of the accretion rate $\lambda_{\rm Edd} \equiv L_{\rm bol}/L_{\rm Edd}$, the nuclear X-ray luminosity $L_{\rm X}$ considered here as a limit for the emission of the disk coronae, and $P_{\rm j}$, expressing the latter two parameters either in the Eddington units, or in the units of the disk luminosity. We find that (i) the accretion rate $\lambda_{\rm Edd}$ in our sample is distributed within a narrow range $\lambda_{\rm Edd} \sim 0.01 - 0.2$; (ii) the normalized jet power $P_{\rm j}/L_{\rm Edd}$ formally correlates with the accretion rate $\lambda_{\rm Edd}$, with some saturation at the largest values $\lambda_{\rm Edd}> 0.05$; (iii) the jet production efficiency $\eta_{\rm jet} \equiv P_{\rm j}/\dot{M}_{\rm acc} c^2$ spans a range from $\eta_{\rm jet} \lesssim 10^{-3}$ up to $\sim 0.2$ at maximum, which is below the level expected for magnetically arrested disks around maximally spinning black holes; and (iv) there is a diversification in $\eta_{\rm jet}$ on the hardness--intensity diagram $L_{\rm X}/L_{\rm bol} - \lambda_{\rm Edd}$, with the jets being produced most efficiently during the high/hard states, and suppressed during the soft states.
\end{abstract}

\keywords{radiation mechanisms: non-thermal --- galaxies: active --- galaxies: jets --- quasars: emission lines --- radio continuum: galaxies --- X-rays: galaxies}

\section{Introduction}
\label{sec:intro}

Compact radio galaxies typically exhibit concave radio continua, with energy flux spectral densities peaking within a narrow frequency range 0.5--10\,GHz; this is the basis for their classification as `GHz-peaked spectrum' (GPS) sources. A subset of them, called `Compact Symmetric Objects' (CSOs), when imaged with high-resolution radio interferometers, display in addition axisymmetric radio morphologies consisting of two lobes at the opposite sides of a radio core, resembling powerful and extended Fanaroff-Riley type II radio galaxies (FR\,IIs, or `classical doubles'), albeit with linear scales $<1$\,kpc \citep[see the review by][]{Odea98}. For those with long-term radio monitoring, one can measure the expansion velocities of the jet terminal hotspots from the nuclei, deriving in this way \emph{kinematic ages} of the radio structures, which are considered as a more reliable proxy for the elapsed time since the jet formation than the ages derived from the sychrotron ageing analysis \citep[see, e.g.,][and references therein]{Orienti16}.

The observed similarity in the radio morphology of CSOs and extended FR\,II radio galaxies, together with young ages of the former class of sources (following either from the kinematic or synchrotron ageing methods), seems to indicate that CSOs represent the phase of newly born jets, which will evolve forming extended structures on $>$\,Myr timescales. However, statistical studies of the corresponding luminosity functions show overabundance of compact radio galaxies over the model expectations assuming a simple evolutionary scenarios \citep{Readhead96,Odea97}. This incoherence may be mitigated if CSOs represent a peculiar class of short-lived radio sources, due to an intermittent jet activity of the central engines \citep{Reynolds97}. Such an intermittency may be caused by a radiation pressure instability within an accretion disk, for example \citep{Czerny09}, although this particular model requires relatively high accretion rates, characteristic for quasar sources rather than low-power radio galaxies. 

This leads to the general yet seldom asked question on the accretion rates in young radio galaxies \citep[see, e.g.,][]{Wu09,Siemiginowska09,Son12}, a question which is not easy to address due to several reasons. The main difficulty here is that the identification of young radio galaxies is based on the broad-band radio spectral and morphological analysis, and not --- as in the case of, e.g., quasars --- any systematic spectroscopic classification in the optical domain. And indeed, optical spectra of compact radio sources are rather diverse, including quasar-like continua with prominent broad emission lines, but also low-ionization or heavily absorbed type spectra. Note that, for example, CSOs are those young radio galaxies which are believed to be observed at large inclinations to the jet axis, and hence in their case one should expect to see rather heavily absorbed nuclear emission, if a circumnuclear dust is already settled in the form of obscuring torii as in the evolved luminous radio galaxies and quasars \citep[see in this context the discussions in][]{Ostorero10,Ostorero17,Willett10,Kosmaczewski19,Sobolewska19b}.

The other general issue is how exactly accretion properties of young radio galaxies shape the launching conditions and the duty cycle of relativistic jets. Here the crucial problem is a robust determination of the total kinetic powers of the newly born jets in such systems, which, again, is hampered by the fact that the radio emission continua of compact jets and lobes are heavily absorbed, due to the synchrotron self-absorption process and/or free-free absorption by the surrounding thermal medium \citep[e.g.,][and references therein]{Callingham15}. Also, the radiative efficiency of compact jets and lobes may be different than that characterizing their evolved analogs, due to the very different environment they reside in and interact with (interstellar vs. intragalactic medium). As a result, all the scaling relation between radio luminosities and jet kinetic powers discussed in the context of extended radio galaxies \citep[e.g.,][]{Willott99,Godfrey16}, should be treated with a particular caution when applied to a sample of young radio galaxies.

On the other hand, with a robust characterization of both the jet kinetic power $P_{\rm j}$ and the accretion-related bolometric luminosity $L_{\rm bol}$ for a well-defined sample of young radio galaxies, along with the central supermassive black hole (SMBH) mass estimates for the analyzed objects, one could investigate the location of the newly-born jet sources on the $P_{\rm j}-L_{\rm bol}$ diagnostic planes. Such diagnostic planes, where different types of active galactic nuclei (AGN) are typically occupying separate loci, are particularly useful when discussing the jet launching processes in general, and possibly also the evolution of such processes during the radio-loud phase of the AGN activity. What follows from the analysis presented in this context in the literature so far, is that the jet production efficiency $P_{\rm j}/L_{\rm bol}$, which can be significantly different for various types of AGN, depends not only on the accretion rate $\dot{m}_{\rm acc} \propto L_{\rm bol}/L_{\rm Edd}$, where $L_{\rm Edd}$ is the corresponding Eddington luminosity, but may also be related to the morphology of the host galaxy either via the entire accretion history over cosmological timescales, shaping the spin distribution of the central SMBHs \citep[and references therein]{Sikora07}, or due to particular accretion conditions determining magnetic structures of the accretion disks in various types of the systems \citep{SB13}.

In this paper, we perform the first systematic study of the accretion rates and the jet production efficiency of young radio galaxies. Our sample selection is based on the published lists of the objects classified robustly as CSOs \citep{Polatidis03,Giroletti09,An12a,An12b,Orienti14,Rastello16}. We further restrict the analysis to the sample of sources for which (i) the redshifts are determined, (ii) the kinematic ages has been measured, and (iii) the core X-ray fluxes could be estimated with high-angular resolution X-ray telescopes ({\it Chandra} or XMM-{\it Newton}). There are 17 sources fulfilling the above criteria, 16 of which are listed in \citet{Siemiginowska16}, including a few with the updated X-ray modeling as presented in \citet{Sobolewska19a,Sobolewska19b}, plus 1323+321 with the kinematic age estimated given in \citet{An12a}, and the X-ray luminosity in \citet{Tengstrand09}. Even though the analyzed sample is rather small, it is composed of the sources which are considered as the most robust and unambiguous examples of \emph{the youngest} radio galaxies, with linear sizes $\lesssim 300$\,pc and corresponding ages $\lesssim 3,000$\,yr (see Table\,1 in \citealt{Siemiginowska16} and also in \citealt{An12a}). 

\begin{deluxetable}{cccc}
\tablecaption{The measured velocity dispersion, and the narrow $H\beta$ line fluxes, for the sources with the available SDSS spectra.\label{table:sdss}}
\tablewidth{0pt}
\tablehead{\colhead{source} & \colhead{$\sigma_\star$ [km\,s$^{-1}$]} & \colhead{$F_{H\beta}$ [erg\,s$^{-1}$\,cm$^{-2}$]} & \colhead{comments}}
\startdata
1031+567  & 218 & 2.01$e-$16 & Type-2 AGN\\
1404+286  & 260 & 4.85$e-$17 & Type-1 AGN\\
1511+0518  & 200 & 8.33$e-$17 & Type-1 AGN\\
1607+26 & 255 & 1.39$e-$15 & Type-2 AGN
\enddata
\end{deluxetable}

For those sources, we gather the available optical spectroscopic data, which allow us to derive the black hole masses and the accretion-related bolometric luminosities (\S\,\ref{sec:optical}); based on the radio data, we estimate the jet kinetic luminosities for the sample, by means of both the established scaling relations, and also anticipating the minimum power condition (\S\,\ref{sec:radio}). The results of the analysis, and in particular a comparison of the accretion rates and the jet production efficiency of the analyzed CSOs with those characterizing the evolved radio-loud AGN, are presented in \S\,\ref{sec:results}, and the main conclusions are summarized in the final \S\,\ref{sec:conclusions}. Throughout the paper we assume $\Lambda$CDM cosmology with the parameters $H_{0}=70$\,km\,s$^{-1}$\,Mpc$^{-1}$, $\Omega _{\rm m}=0.3$, and $\Omega _{\Lambda}=0.7$.

\section{Optical Spectral Information}
\label{sec:optical}

We gathered the optical spectral information available for the CSOs from the list of \citet{Siemiginowska16} plus 1323+321 \citep{An12a}. This includes the Sloan Digital Sky Survey (SDSS) spectra for the four targets, which are analyzed in detail in the following Section \,\ref{sec:SDSS}. For the remaining objects, we have collected the data presented in the literature, as described in Section \,\ref{sec:literature}. Alltogether, we were able to provide the SMBH mass estimate for all but one source, 1843+356, while the bolometric accretion luminosities could not be determined for only three sources, namely 1843+356, 1946+708, and 1943+546. 

\subsection{SDSS Data}
\label{sec:SDSS}

We have searched the SDSS database \citep{sdss} for the available optical information for the sources from our sample, finding only four good-quality spectra with detected absorption and emission lines, as required for our purposes. We used these data to estimate the bolometric luminosities and black hole masses for 1607+26, 1511+0518, 1404+286, and 1031+567.

\begin{figure*}[t]
\centering
\includegraphics[width=1.\columnwidth]{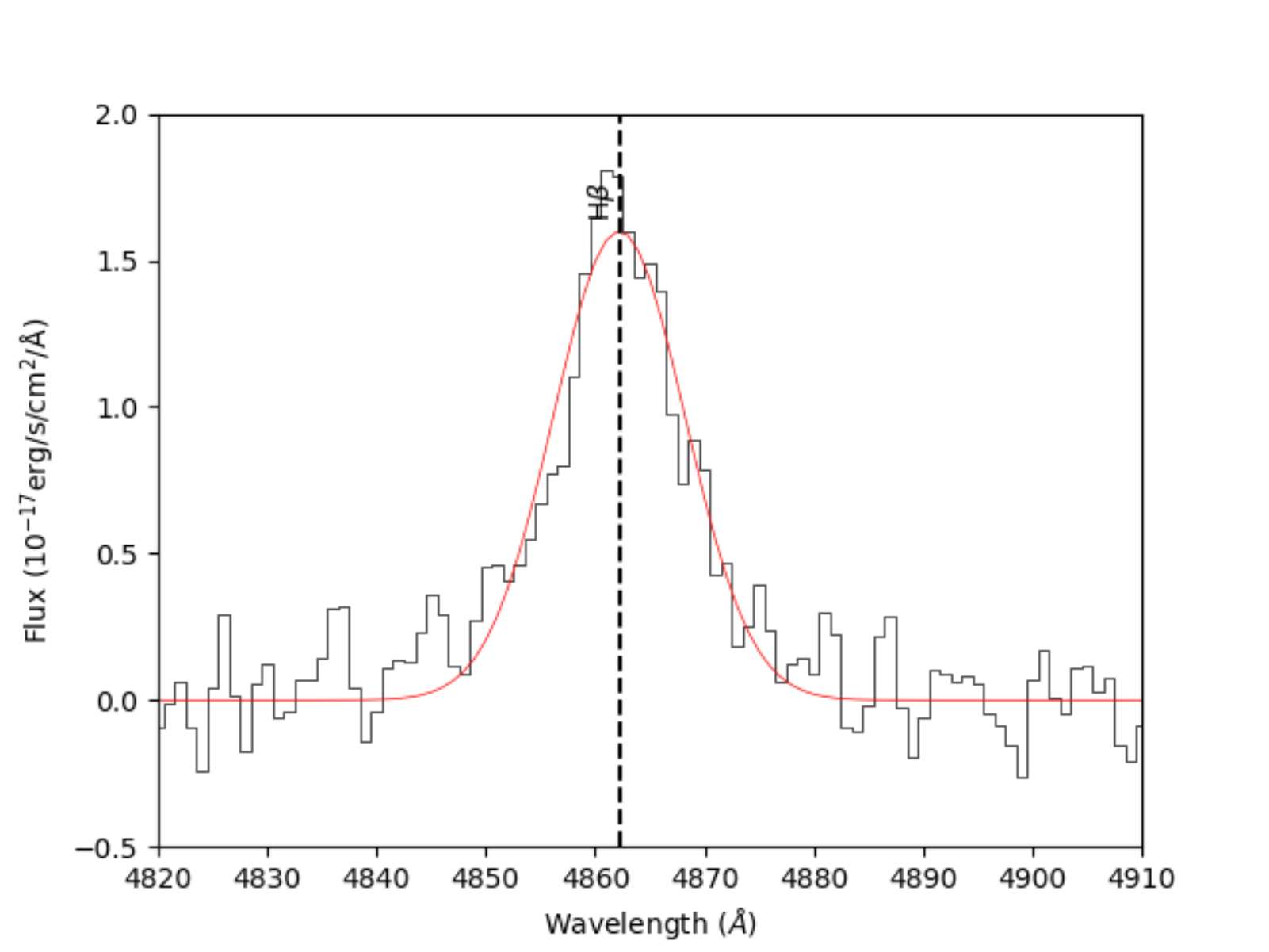}
\includegraphics[width=1.\columnwidth]{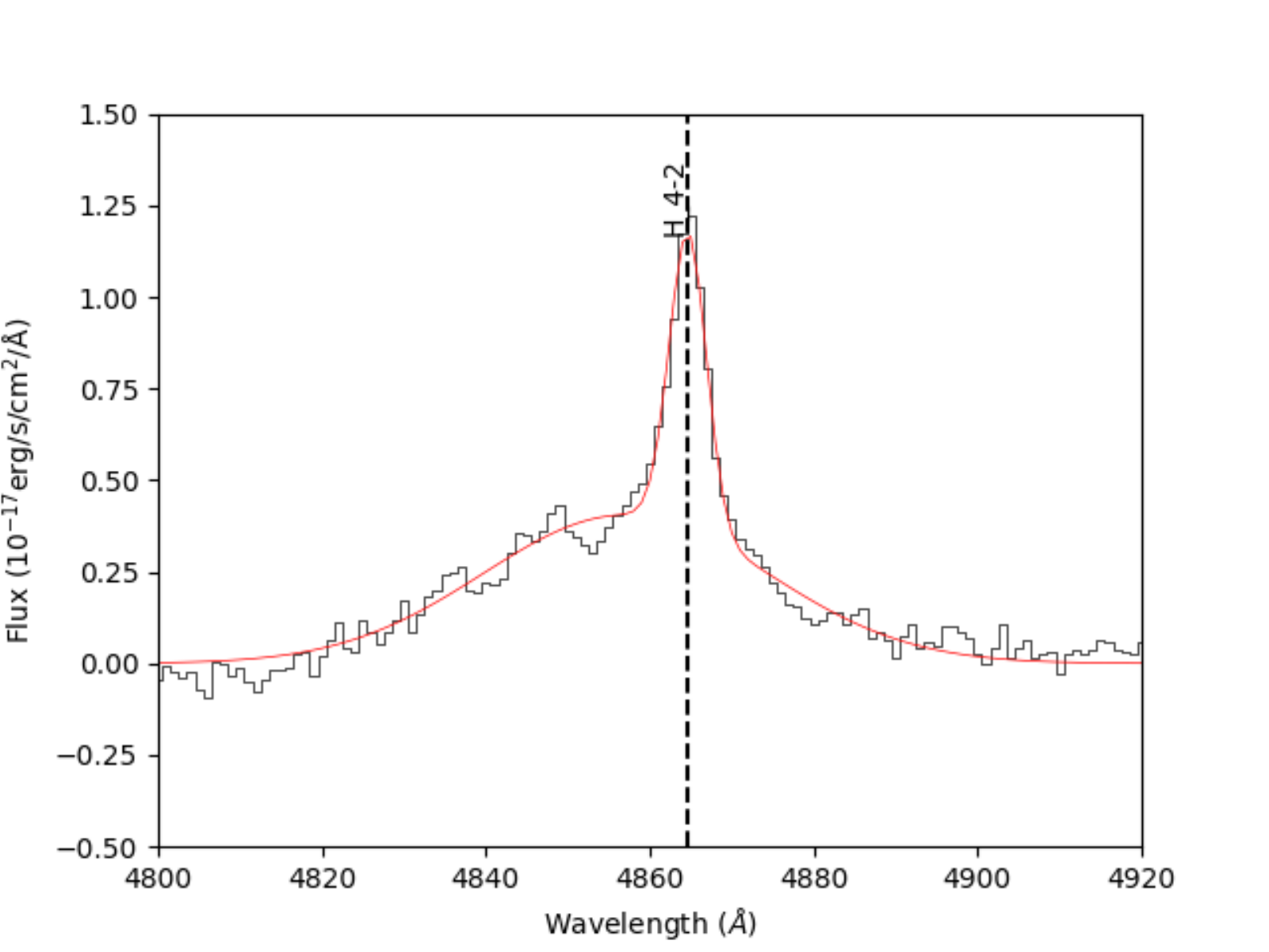}
\includegraphics[width=1.\columnwidth]{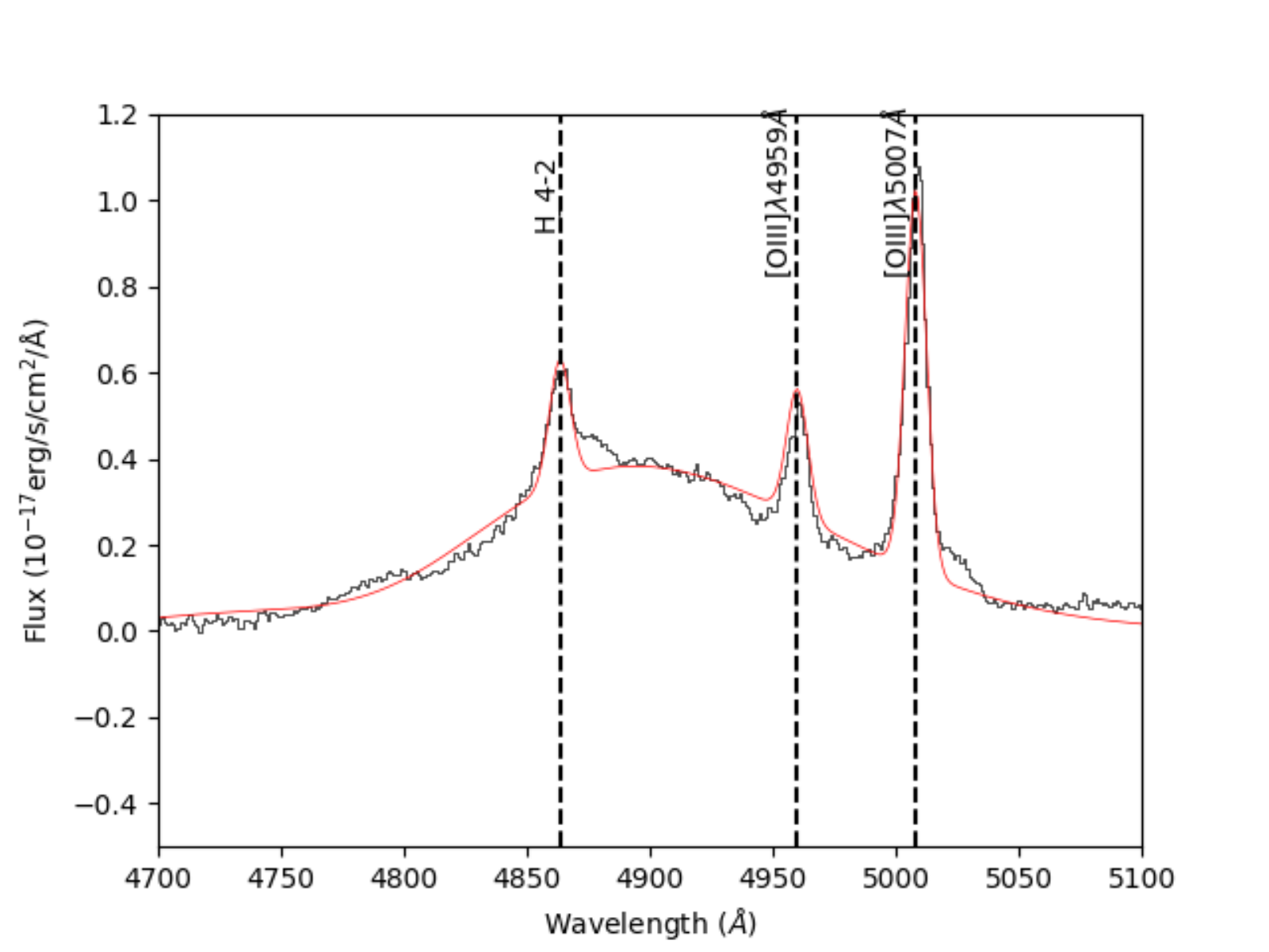}
\includegraphics[width=1.\columnwidth]{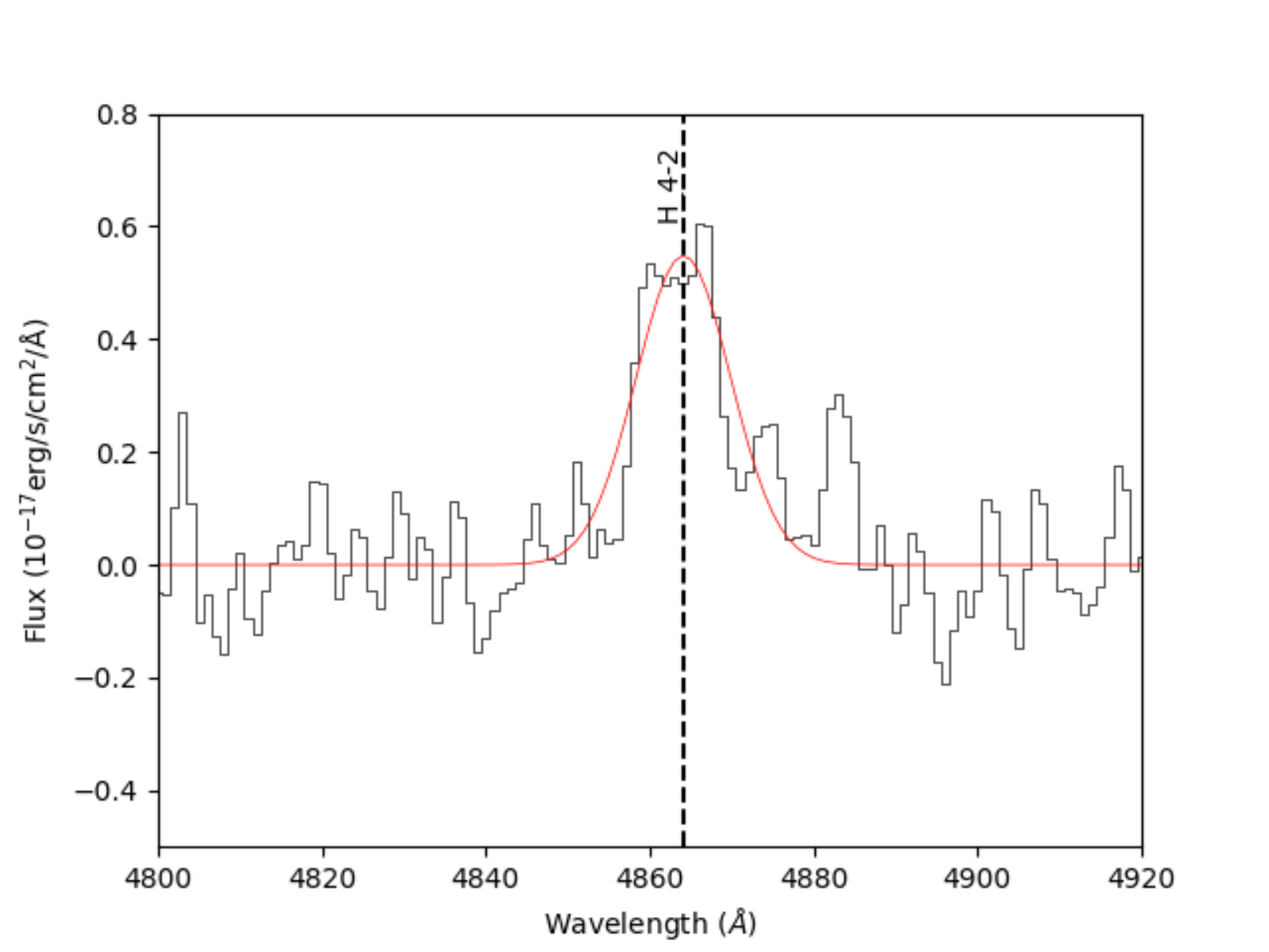}
\caption{Optical spectra around the $H\beta$ emission line (black lines) and best fits to the line profiles (red lines) for 1031+567 (top left), 1404+286 (top right), 1511+0518 (bottom left), and
1607+26 (bottom right).\label{fig:lines}}
\end{figure*}
The spectra were corrected for Galactic extinction (using the maps of \citealt{Schlegel98} and the extinction law of \citealt{Cardelli89}), and cosmological redshift; they were resampled to $\delta\lambda=1\AA$ spacing with the NOAO IRAF\footnote{IRAF is distributed by the National Optical Astronomy Observatory, which is operated by the Association of Universities for Research in Astronomy (AURA) under cooperative agreement with the National Science Foundation.} package. In order to extract the stellar components from the composite starlight $+$ AGN spectra, we used the synthesis code \texttt{STARLIGHT} \citep{cid}. The code fits the continuum and absorption lines as a combinations of single stellar populations extracted from the models of \citet{Bruzual03}. In the fitting procedure, we masked the emission lines, provided the information about flagged regions and errors, and specified manually (in the configuration file) the AGN continua along with the broad emission lines in the cases of the two sources 1404+286 and 1511+0518.

Based on the obtained synthetic starlight spectra, we derived the basic information regarding stellar composition, total stellar masses, and stellar velocity dispersions $\sigma_{\star}$ (corrected for the base 64\,km\,s$^{-1}$ and the instrumental velocity dispersion 70\,km\,s$^{-1}$), as summarized in Table \,\ref{table:sdss}. With such, we estimated the SMBH masses for our targets by using the established $\mathcal{M}_{\rm BH}-\sigma_\star$ scaling relation for ellipticals and bulge-like systems,
\begin{equation}
\log{\left(\frac{\mathcal{M}_{\rm BH}}{M_\odot}\right)}=\alpha+\beta \times \log{\left(\frac{\sigma_{\star}}{200\,{\rm km\,s^{-1}}}\right)} \, ,
\label{eq:MBH}
\end{equation}
with $\alpha = 8.13\pm0.05$ and $\beta = 5.13\pm 0.34$ \citep{Graham11}.

The stellar model components were then subtracted from the corrected SDSS spectra, in order to estimate the bolometric luminosities of the AGN, using the narrow $H\beta$ line. Our choice was motivated by the fact that no significant variation of this line flux with the ionization parameter was noted \citep{Netzer09}. Even hough the $H\alpha$ line is one of the most prominent lines in AGN spectra in general, and as such is frequently used for estimating the $L_{\rm bol}$ value, in the cases of the objects analyzed here it was ruled out from the analysis due to the blending with the [NII] lines. The $H\beta$ lines detected in our residual AGN spectra were fitted with Gaussian profile using the \texttt{Python Spectroscopic Toolkit} \citep{Ginsburg11}, as presented in Figure \,\ref{fig:lines}. For the two sources 1404+286 and 1511+0518, where the broad components of the $H\beta$ line have been detected as well (the blue-shifted broad component in the case of 1511+0518, and both the red- and blue-shifted components in the case of 1404+286), a multi-component Gaussian fitting was performed. Based on the fits, we estimated the $H\beta$ line luminosities $L_{H\beta} = 4 \pi d_{\rm L}^2 F_{H\beta}$ (see Table \,\ref{table:sdss}), and then the corresponding AGN-related bolometric luminosities using the \citet{Netzer09} relation
\begin{equation}
\log L_{\rm bol}=\log L_{H\beta}+3.48 \, .
\label{eq:Netzer}
\end{equation}
We finally corrected the luminosity values for the reddening in host galaxies by requiring $H\alpha/H\beta=2.86$ and assuming the $\lambda^{-0.7}$ reddening law.

In the cases of the type-1 AGN  in our SDSS sample, i.e. 1404+286 and 1511+0518, which curiously are also classified as Compton-thick based on the X-ray spectroscopy \citep{Guainazzi04,Siemiginowska16,Sobolewska19a,Sobolewska19b}, the bolometric luminosities obtained as described above should be considered as not reliable, because they significantly underestimate the true radiative power of the AGN; the major source of the uncertainty here is the problem with extracting relatively weak narrow $H\beta$ fraction from much more prominent broad components. For 1511+0518, we have therefore decided to use the $L_{\rm bol}$ estimate following from the SED fitting by \citet{Trichas13}, while for 1404+286 we used the scaling between the $L_{\rm bol}$ and the 12\,$\mu$m luminosity established for quasar sources by \citet{Richards06}, utilizing the Infrared Astronomical Satellite (IRAS) data \citep[see][]{Kosmaczewski19}.

\begin{deluxetable}{cccc}
\tablecaption{Bolometric luminosities estimated from the narrow $H\beta$ measurements found in the literature. \label{table:Hbeta}}
\tablewidth{0pt}
\tablehead{	
\colhead{source} & \colhead{$H\alpha/H\beta$} & \colhead{correction method} & \colhead{$L_{H\beta}$ [erg\,s$^{-1}$]}}
\startdata
0035+227 & 2.73 & average & 1.98$e+$41 \\
0710+439 & -- & average & 1.54$e+$42 \\
1245+676 & 1.84 & average & 1.36$e+$41 \\
1323+321 & 2.8 & average & 2.07$e+$42 \\
1718--649 & 3.4 & $H\alpha/H\beta$  & 1.25$e+$41 \\
1934--634 &  5 & $H\alpha/H\beta$ & 1.45$e+$41\\
2352+496 & 4.57 & $H\alpha/H\beta$ & 2.65$e+$41
\enddata 
\end{deluxetable}

\subsection{Other Data}
\label{sec:literature}

We have searched the literature for the emission line measurements in the spectra of the remaining 13 sources from our list. For the majority of those, the AGN bolometric luminosities were estimated by \citet{Wu09} from the [OIII] 5007\,$\AA$ luminosities. However, this method may not to be the most reliable one in the case of newly-born jetted AGN. That is because, according to the standard model, the narrow [OIII] line is produced when atoms are photo-ionized by the UV radiation originating in the accretion disk, while in young radio-loud AGN the additional source of the ionization may be provided by the interaction of compact jets with the ISM clouds \citep[see the discussion in][and references therein]{Ostorero10}. For this reason, we have carefully investigated the optical spectra available in the literature, to search for the narrow $H\beta$ line. This resulted in the $L_{H\beta}$ luminosity measurements, performed in a similar manner as described in a previous Section \,\ref{sec:SDSS}, for six sources from the sample, as summarized in Table \,\ref{table:Hbeta} (with the corresponding references provided in Table \,\ref{table:info}). Here we have corrected the measured $H\beta$ luminosities for the reddening assuming the $\lambda^{-0.7}$ law, with the correction factors calculated from either the observed $H\alpha/H_\beta$ ratio, or (if $H\alpha/H\beta<2.86$) by assuming the average value established for Seyfert type\,2 galaxies by \citet{Netzer09}.

This leaves us with six CSOs from the list, for which the bolometric AGN luminosities could not be determined based on the narrow $H\beta$ lines. From those, in the cases of 0108+388, 0116+319, and 2021+614, for the following analysis we adopted the $L_{\rm bol}$ values estimated based on the [OIII] line by \citet{Wu09}, noting however a very low signal-to-noise ratio in the spectra of the former two sources. For the remaining three targets, i.e. 1843+356, 1946+708, and 1943+546, we could not find any appropriate optical spectroscopic data \citep[see in this context][]{Henstock97,Caccianiga02}.

As for the black hole mass estimates for the objects of our sample which were not covered by the SDSS, the $\mathcal{M}_{\rm BH}-\sigma_{\star}$ relation could be applied only for the three sources (0035+227, 1943+546, and 1245+676) for which the stellar velocity dispersion was measured by \citet{Son12}. The remaining masses were adopted from \citet{Willett10} or \citet{Wu09}, who applied the scaling relations involving either the galaxy bulge luministy $L_{\rm blg}$ \citep[see][]{Bentz09}, or the absolute R-band optical magnitude of the host $M_{\rm R}$ \citep[see][]{McLure04}, respectively, as summarized in Table \,\ref{table:info}. The $\mathcal{M}_{\rm BH}$ value could not be estimated only in the case of 1843+356.

\section{Jet Kinetic Luminosities}
\label{sec:radio}

Total radio fluxes of CSOs are not expected to be characterized by any strong variability of the colored-noise type, as seen in the sources for which the radiative output is dominated by the beamed jet component \citep[i.e., blazars; see, e.g.,][and references therein]{Goyal17}. Still, they may exhibit smooth monotonic flux changes on timescales of years and decades, due to (i) the fast expansion of their compact lobes, and (ii) the spectral evolution around the peak frequencies, due to the changes in the absorption properties of the medium within and outside the lobes. Both these effects have to be kept in mind when estimating the jet kinetic luminosities of the youngest radio galaxies from their monochromatic radio fluxes.

In our first approach, we derive the jet kinetic power for the objects from our list, $P_{\rm j}$, by applying the calorimetric scaling relation established for the evolved radio sources by  \citet{Willott99}, as discussed by \citet[equation 2 therein]{Rusinek17}, namely
\begin{equation}
P_{{\rm j}\,(W)}= 5.0 \times 10^{22} \, \left(\frac{L_{\rm 1.4\,GHz}}{\rm W\,Hz^{-1}}\right)^{6/7} {\rm erg\,s^{-1}}\, .
\label{eq:Will}
\end{equation}
The corresponding 1.4\,GHz luminosity spectral densities $L_{\rm 1.4\,GHz}$ for the majority of the analyzed sources were provided by the NRAO VLA Sky Survey \citep[NVSS;][]{Condon98}; in the cases of the two southern sources from the list (1718--649 and 1934--638), we used the ATCA Monitoring Observations \citep{Tingay03}. These are all listed in Table \,\ref{table:info}.  We note that the $P_{\rm j} - L_{\rm 1.4\,GHz}$ relation was lately updated in \citet{Godfrey13} separately for FR\,I and FR\,II sources; with the new scalling provided by the authors for classical doubles, the jet powers derived for the sample considered here turn out to be lower by a factor of a few when compared to the values emerging using equation \ref{eq:Will} above.

In our second approach, we derive the \emph{minimum} jet powers, by assuming energy equipartition between magnetic field and radio-emitting electrons within the lobes, and utilizing the fact that for all the sources studied here, the kinematic ages $\tau_{\rm j}$ have been measured. In particular, we adopt the $\tau_{\rm j}$ values as listed in Table \,\ref{table:info}, along with the corresponding references; these are all consistent with the values adopted by \citet{Siemiginowska16} and listed in \citet{An12a}, except for 1031+567, for which we take 620\,yr following \citet{Taylor00}. With such, we find the jet kinetic luminosity as $P_{\rm j}=\mathcal{H}_{\rm j}/ \tau_{\rm j}$, where $\mathcal{H}_{\rm j} = 4 p_{\rm min} V$ is the source enthalpy corresponding to the lobes' minimum pressure $p_{\rm min}$ and the volume $V$. 

The youngest radio galaxies are expected to be characterized by relatively large aspect ratios of their lobes, with the minor-to-major semi-axis ratios $b/a \approx 0.25$ \citep{Kawakatu08}. For the given linear sizes of the analyzed sources, $LS = 2 a$, all listed in Table \,\ref{table:info}, we therefore introduce the effective radius of the lobes as $R_{\rm eff} = \sqrt[3]{\frac{3}{4} a b^2} \simeq 0.18 \times LS$. With such, we derive the equipartition magnetic field within the lobes with the ''classical'' formula, $B_{\rm eq}=[4.5\, c_{12}\, L_{\rm rad}]^{2/7}\, R_{\rm eff}^{-6/7}$\,G, where $c_{12} \simeq 3\times 10^7$ (in cgs units), and $L_{\rm rad}$ stands for the total radio power emitted in the frequency range 0.01--100\,GHz \citep[see][]{Beck05}. This power is calculated here from the observed monochromatic 5\,GHz luminosities (see Table \,\ref{table:info}), assuming that the \emph{intrinsic} radio continua in all the cases could well be approximated by a simple power-law with mean spectral index $\alpha=0.73$ above the peak frequency, as derived for young radio galaxies by \citet{deVries97}; this yields $L_{\rm rad} \simeq 7.62 \times L_{\rm 5\,GHz}$. Finally, for the minimum pressure we take $p_{\rm min} = 2 \times p_B$, where the magnetic pressure is $p_B=B_{\rm eq}^2/8\pi$, obtaining 
\begin{eqnarray}
P_{\rm j} && = \frac{4 B_{eq}^2 R_{\rm eff}^3}{3 \, \tau_{\rm j}} \sim 1.5 \times 10^{45} \times \\ && \left(\frac{LS}{\rm 100\,pc}\right)^{9/7} \, \left(\frac{\tau_{\rm j}}{\rm 100\,yr}\right)^{-1} \, \left(\frac{L_{\rm 5\,GHz}}{\rm 10^{42}\,erg/s}\right)^{4/7} \, {\rm erg\,s^{-1}} \, .\nonumber  \label{eq:min} 
\end{eqnarray}

\begin{figure}
\centering
\includegraphics[width=1.0\columnwidth]{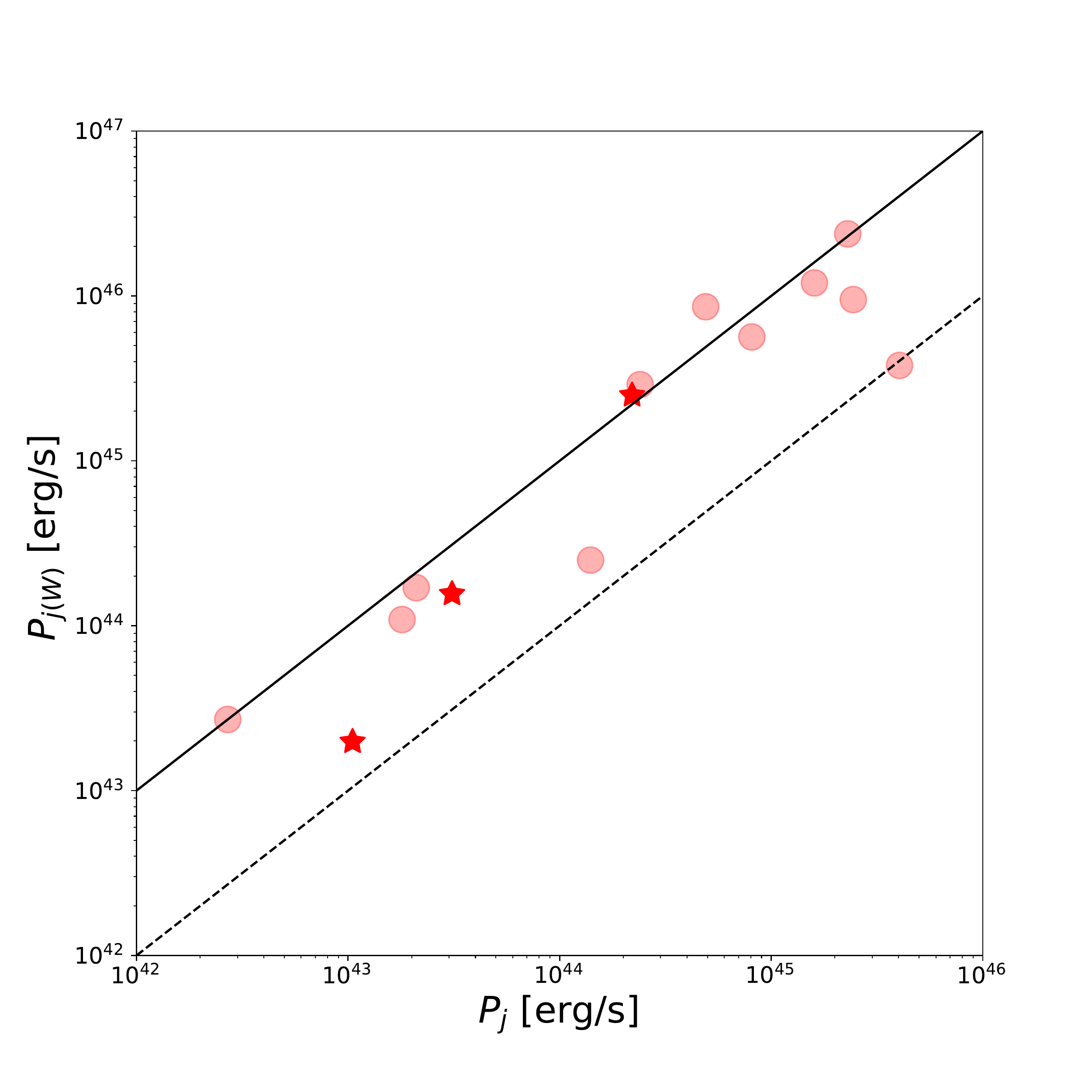}
\caption{The jet power $P_{{\rm j}\,(W)}$ of the CSOs of our sample as derived from the \citet{Willott99} scaling relation, as a function of the minimum jet kinetic luminosities, $P_{\rm j}$. The sources confirmed as Compton-thick (2021+614, 1511+0518, and 1404+286) are denoted by red stars. Black dashed and solid lines illustrate the $P_{{\rm j}\,(W)} = P_{\rm j}$ and $P_{{\rm j}\,(W)} = 10 \, P_{\rm j}$ scalings, respectively. \label{fig:radio}}
\end{figure}

The resulting estimates are listed in Table \,\ref{table:info}, and plotted against $P_{{\rm j}\,(W)}$ in Figure \,\ref{fig:radio}. As shown, in the majority of the cases, the minimum jet kinetic luminosities are lower by one order of magnitude than the corresponding jet powers derived from the \citet{Willott99} scaling relation  (and so by a factor of a few than the corresponding jet powers derived from the \citealt{Godfrey13} scaling relation). This is not surprising, as on the one hand, the $P_{\rm j}$ values provide (by definition) only safe lower limits for the jet power, while on the other hand, the $P_{{\rm j}\,(W)}$ values obtained from the radio flux scaling calibrated for the evolved radio galaxies, are expected to over-estimate the jet power in young radio sources, due to the anticipated enhanced radiative efficiency of compact radio-emitting jets and lobes interacting directly with the ISM of host galaxies \citep[see the discussion in][]{Tadhunter11,Dicken12}.

Interestingly, one can compare our minimum power estimates with the jet kinetic luminosities derived in \citet{Ostorero10} by means of modeling of the broad-band spectral energy distributions of the selected GPS/CSO sources. There are five objects overlapping between the two samples, namely 0108+388, 0710+439, 1031+567, 1404+286, and 2352+495. For all of those but 1404+286, our values $P_{\rm j}$ agree very well with those derived by \citeauthor{Ostorero10} (see Table \,3 therein), being in particular lower by a small factor ranging from 0.9 down to 0.4. Only in the case of 1404+286, however, the discrepancy is as large as one order of magnitude. And indeed, the radio continuum of this very compact source is heavily absorbed at longer wavelengths, what manifests in the observed high peak frequency. For this reason, the jet kinetic luminosity for this target may possibly be under-estimated here.

\citet{Ostorero10} assumed that the bulk of the observed X-ray emission of the analyzed targets is due to non-thermal radiative output of compact lobes \citep[see the model description in][]{Stawarz08}. In general, the origin of the X-ray emission in young radio galaxies is still under debate, and the accretion disk coronae, radio jets and lobes, as well as the hot gaseous fraction of the ISM of host galaxies, are all viable options \citep[see the discussion in][]{Siemiginowska08,Siemiginowska16,Tengstrand09,Kunert14,Kosmaczewski19}. Only in the case of targets confirmed as ``Compton-thick'', the coronal scenario should be considered as the most probable one \citep{Sobolewska19a,Sobolewska19b}. Yet by considering that the dominant fraction of the observed X-ray flux is produced in compact lobes, \citeauthor{Ostorero10} derived strict upper limits for the corresponding jet powers. The fact that those upper limits are higher than our lower limit estimates by only a small factor up to two (except of 1404+286, the only Compton-thick source in the overlapping sub-sample), implies that CSOs are in general close to the minimum-power condition indeed.

Keeping in mind the above-mentioned ambiguity regarding the origin of the X-ray emission of young radio galaxies, we consider the 2--10\,keV luminosities for the analyzed CSOs (corrected for the Galactic and internal absorption, as summarized in Table \,\ref{table:info} along with the corresponding references), as a proxy for the radiative output of the accretion disk coronae. 

\begin{deluxetable}{ccccc}
\tablecaption{Core radio fluxes and monochromatic luminosities. \label{table:cores}}
\tablewidth{0pt}
\tablehead{	
\colhead{source} & \colhead{$\nu$}\,[Hz] & \colhead{$S_{\rm c}$\,[mJy]} & \colhead{$L_{\rm c}$\,[erg\,s$^{-1}$]} & \colhead{ref.}}
\startdata
0035+227  &	4.3 & 18$^\dagger$ & 1.6E+40 & (1)\\
0108+388 &	15 & 10  & 2.7E+42 & (2)\\
0116+319 &	5 & 24  & 9.3E+39 & (3)\\
0710+439 &	15 & 45 & 6.6E+42 & (2)\\
1323+321 &    4.9 & 9.1 & 2.2E+41 & (4,5)\\
1511+0518 &	8.4 & 18.5 & 2.5E+40 & (6)\\
1943+546 &	5 & 24 &  2.4E+41 & (7)\\
1946+708 &	 8.4 & 90$^\ddagger$ & 1.8E+41 & (8)\\
2352+495 &	15 & 12 & 2.8E+41 & (2)
\enddata 
\tablecomments{references (1) \citet{Polatidis08}, (2) \citet{Taylor96}, (3) \citet{Giroletti03}, (4) \citet{Helmboldt07}, (5) \citet{Tremblay16}, (6) \citet{An12b}, (7) \citet{Xu95}, (8) \citet{Taylor09};\\
$^\dagger$ measured peak flux in VLBA 4.3\,GHz map from August 2013 in RFC database \citep{Petrov19}; $^\ddagger$ maximum core flux.}
\end{deluxetable}

\begin{figure}
\centering
\includegraphics[width=1.15\columnwidth]{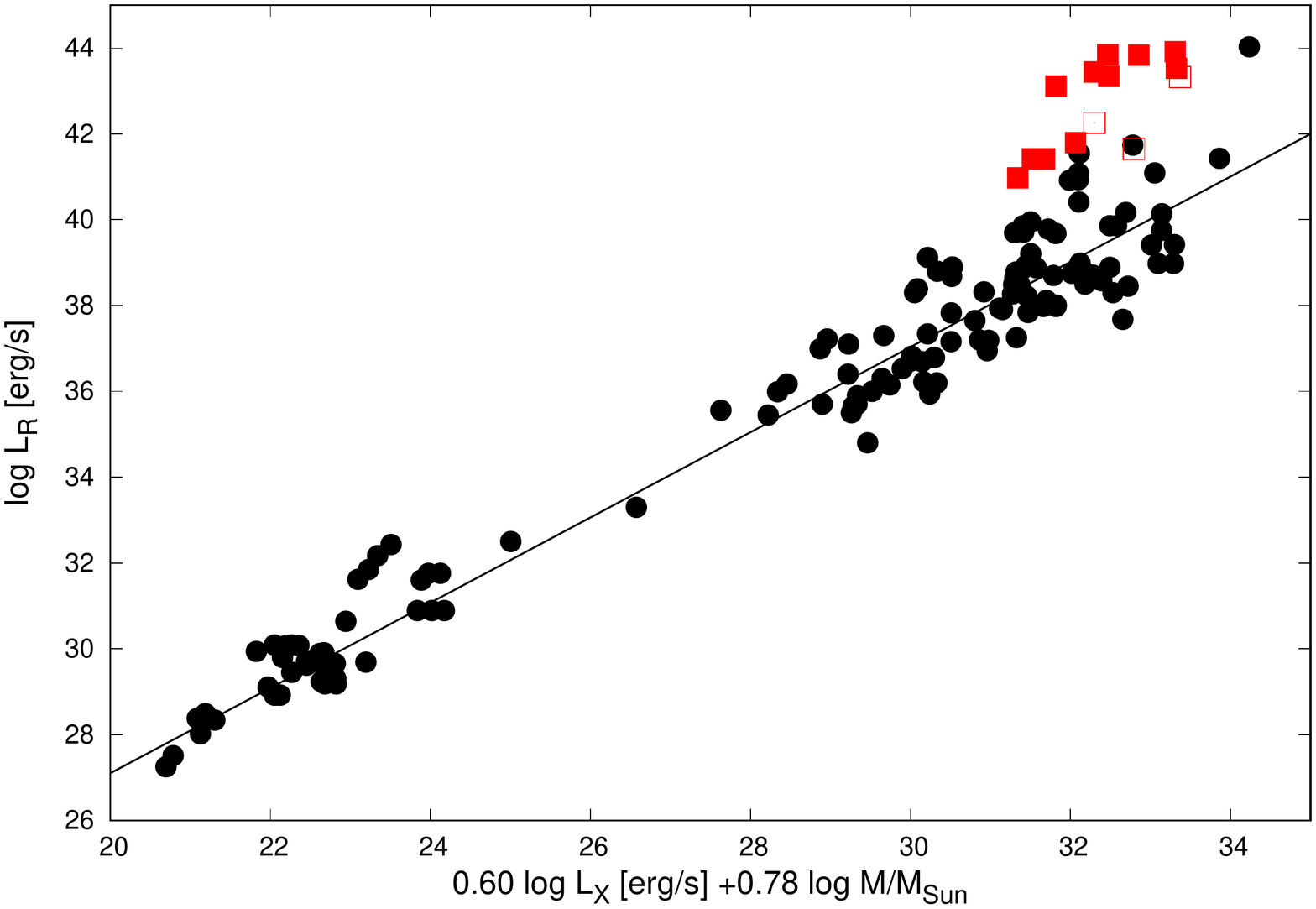}
\includegraphics[width=1.15\columnwidth]{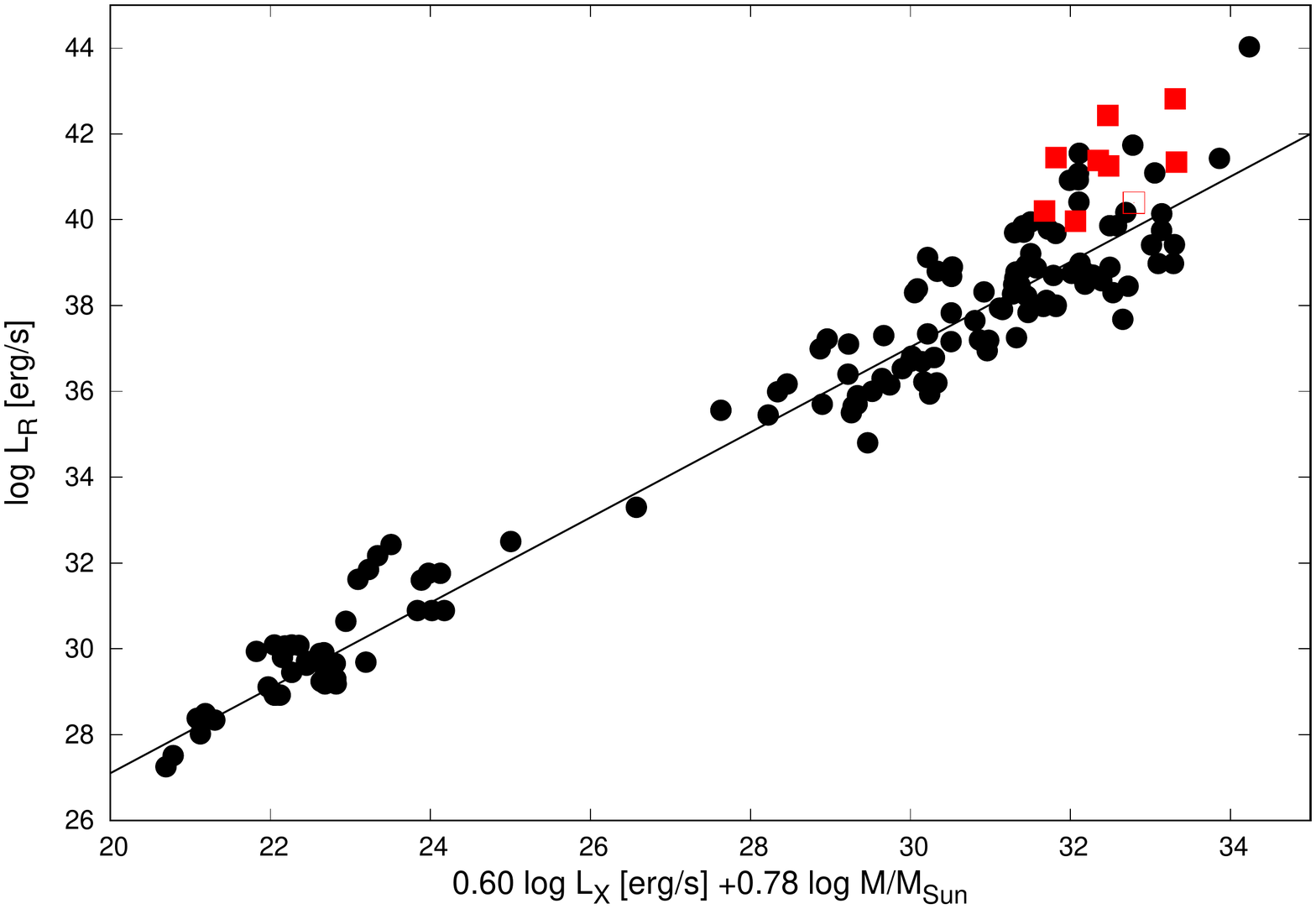}
\caption{{\it Upper panel:} Distribution of the analyzed sample of CSOs in the Fundamental Plane for the black hole activity, as introduced by \citet{Merloni03}. Our sources, for which the \emph{total} radio (5\,GHz) and X-ray (2--10\,keV) luminosities are used, are marked by red squares (filled for Compton-thin sources, empty for Compton-thick objects); sources included in the \citeauthor{Merloni03} sample, i.e. AGN and X-ray Binaries with measured core luminosities, are marked in the figure by black circles, and the corresponding best-fit regression line is shown as a black solid line. {\it Lower panel:} same as in the upper panel, except that here for the analyzed CSOs the monochromatic \emph{core} radio fluxes are used. \label{fig:FP}}
\end{figure}

In this context, let us comment on the distribution of the sample in the so-called Fundamental Plane for the black hole activity, as introduced by \citet{Merloni03}, utilizing first the \emph{total} 5\,GHz luminosities of the analyzed sources (as listed in Table \,\ref{table:info}). As shown in the upper panel of Figure \,\ref{fig:FP}, our CSOs (marked by filled red squares, or empty red squares in the case of the Compton-thick objects), are located above the best-fit correlation line, being in particular over-luminous in radio with respect to the sources (AGN and X-ray Binaries) included in the \citeauthor{Merloni03} sample (and marked in the figure by filled, black circles). This effect may however be explained by a dominant contribution of the lobes' radiative output to the observed total radio fluxes of the considered CSOs (in an analogous way as speculated by \citealt{Saikia18} for their SDSS-FIRST sample of AGN; see also the discussion in \citealt{Fan16}). Indeed, if the Fundamental Plane does reflects some fundamental property of the accreting black holes in the astrophysical environment, these are strictly the core radio (and X-ray) fluxes which should be utilized in the analysis. 

The problem is that, in the general case of CSOs, radio cores are often undetected \citep[e.g., 1404+286; see][]{Luo07,Wu13}, or detected only at higher radio frequencies. When detected, the compact cores of CSOs constitute typically a few percents of the total radio fluxes. We have reviewed the existing VLBI literature and, whenever necessary, inspected the available radio maps, to search for the core detections for the objects in our sample. Altogether, excluding the cases with the uncertain core identification, we were able to provide core flux measurements at frequencies ranging from 4.3\,GHz to 15\,GHz for the nine targets, as listed in Table \,\ref{table:cores}. With such, the analyzed CSOs are more in agreement with the Fundamental Plane by \citet{Merloni03}, although are still above the best-fit correlation line, as shown in the lower panel of figure \,\ref{fig:FP}.

\section{Jet Production Efficiency Parameters}	
\label{sec:results}

Here we investigate the distribution of our CSO sample in the three-dimensional space of the jet kinetic luminosity $P_{\rm j}$, the bolometric luminosity of the accretion disk $L_{\rm bol}$, and the X-ray luminosity $L_{\rm X}$. As anticipated in Section \,\ref{sec:radio}, here $L_{\rm X}$ is considered as a proxy for the radiative output of the accretion disk coronae.
\begin{figure}
\centering
\includegraphics[width=0.8\columnwidth]{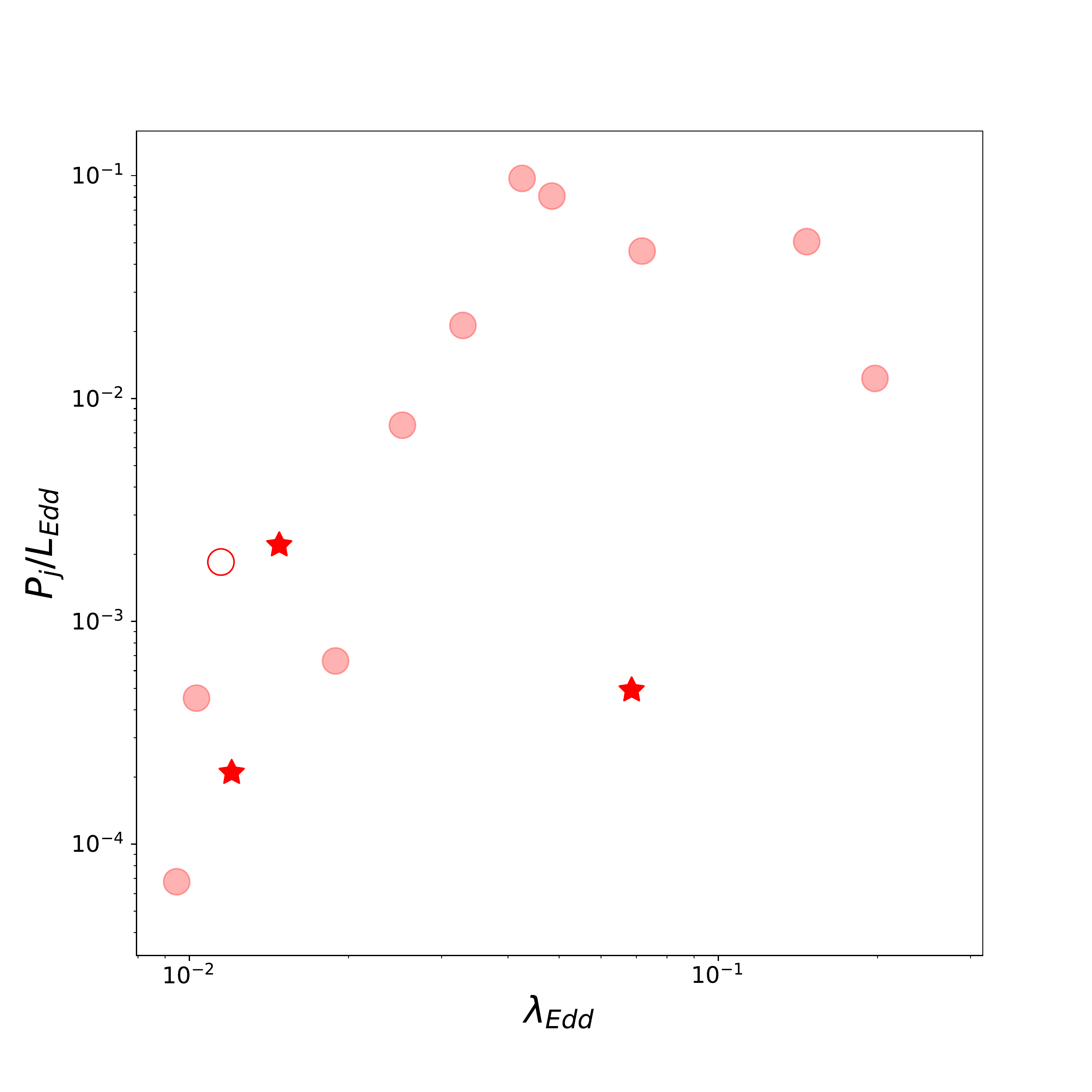}\\
\includegraphics[width=0.8\columnwidth]{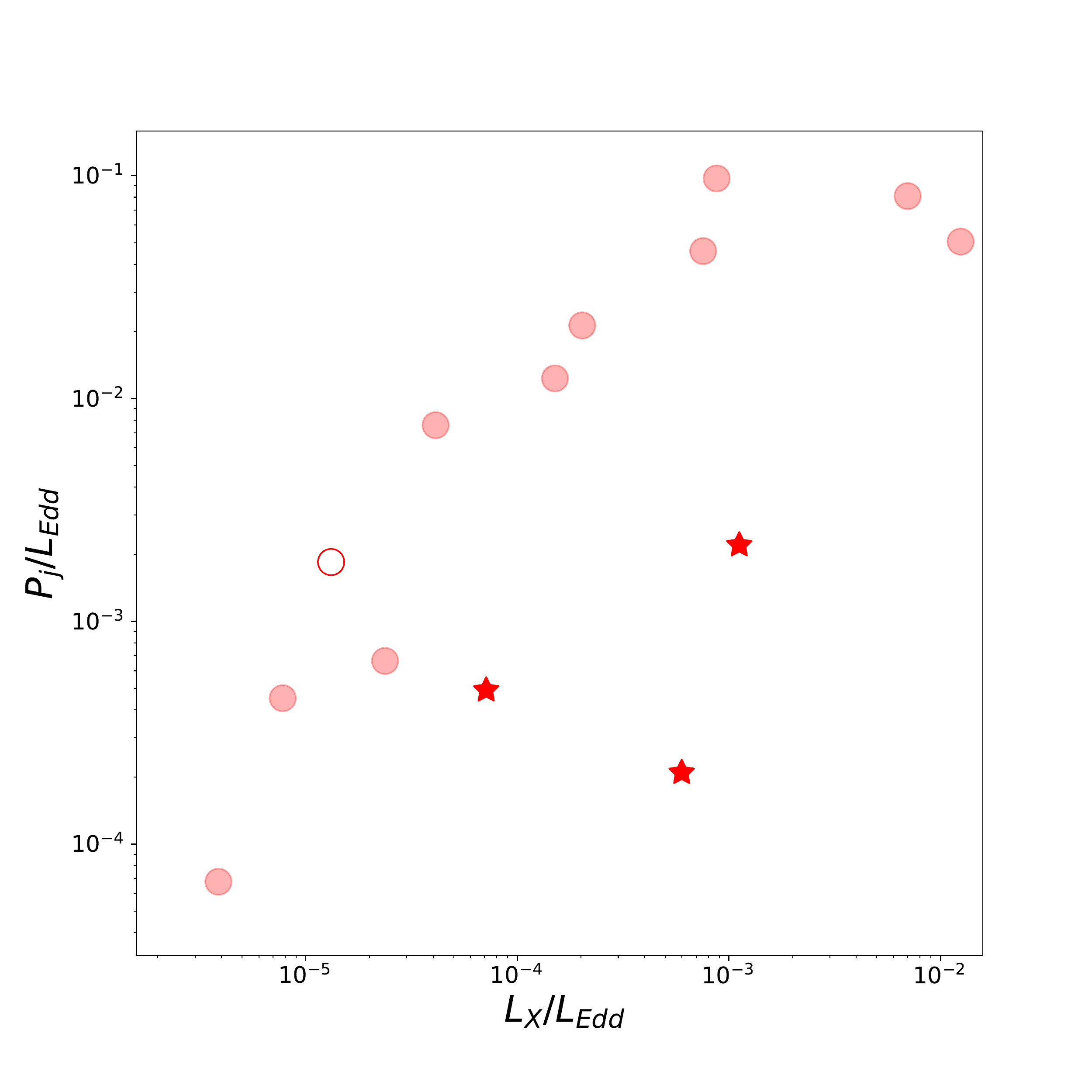}\\
\includegraphics[width=0.8\columnwidth]{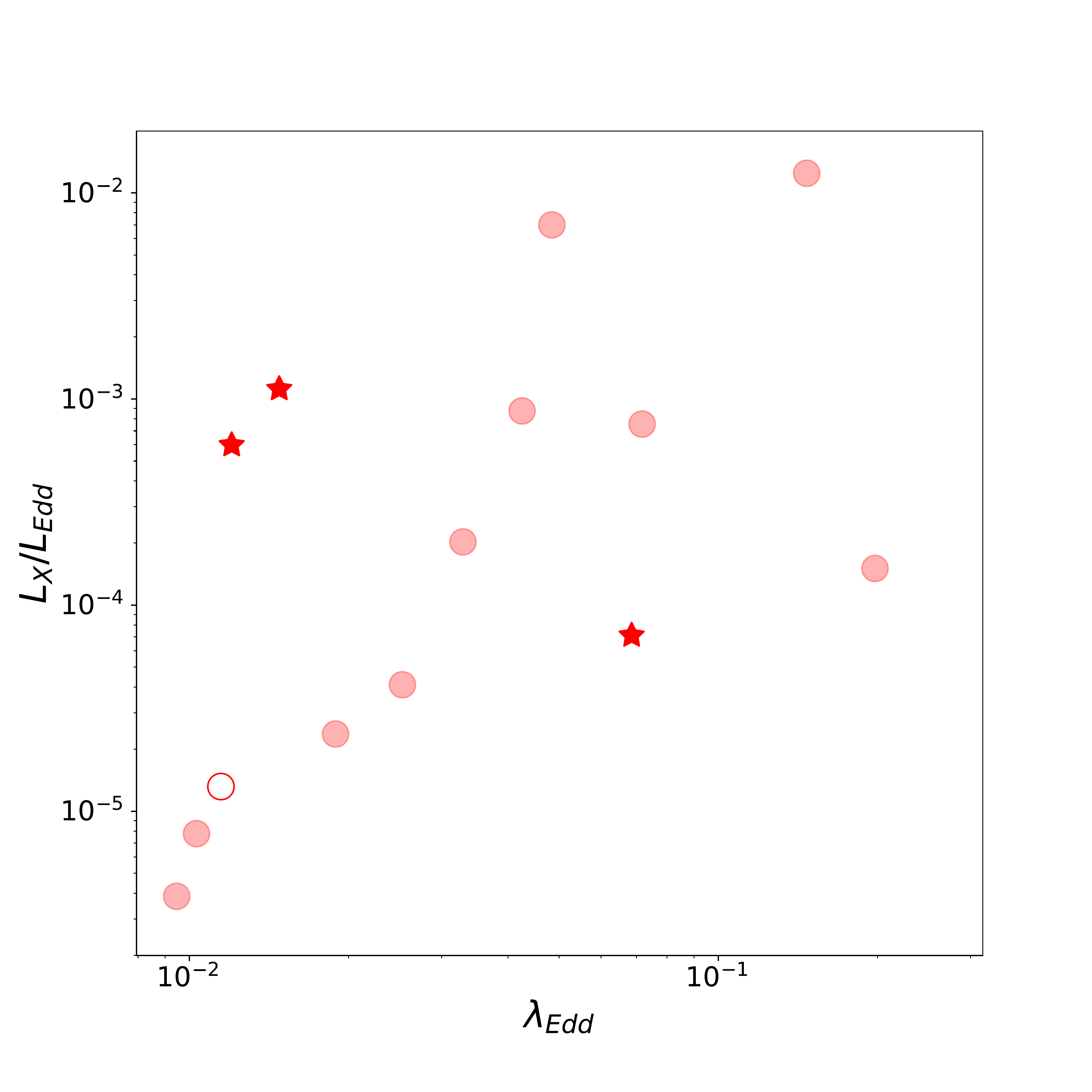}
\caption{Distribution of our CSO sample in the three-dimensional space of the jet kinetic luminosity in Eddington units, $P_{\rm j}/L_{\rm Edd}$, the accretion rate $\lambda_{\rm Edd} \equiv L_{\rm bol}/L_{\rm Edd}$, and the X-ray luminosity in Eddington units, $L_{\rm X}/L_{\rm Edd}$. The empty red circle marks 0116+319, with the upper limit for its X-ray luminosity, while the sources confirmed as Compton-thick, namely 2021+614, 1511+0518, and 1404+286, are denoted by red stars. \label{fig:Edd}}
\end{figure}

\begin{figure}
\centering
\includegraphics[width=0.8\columnwidth]{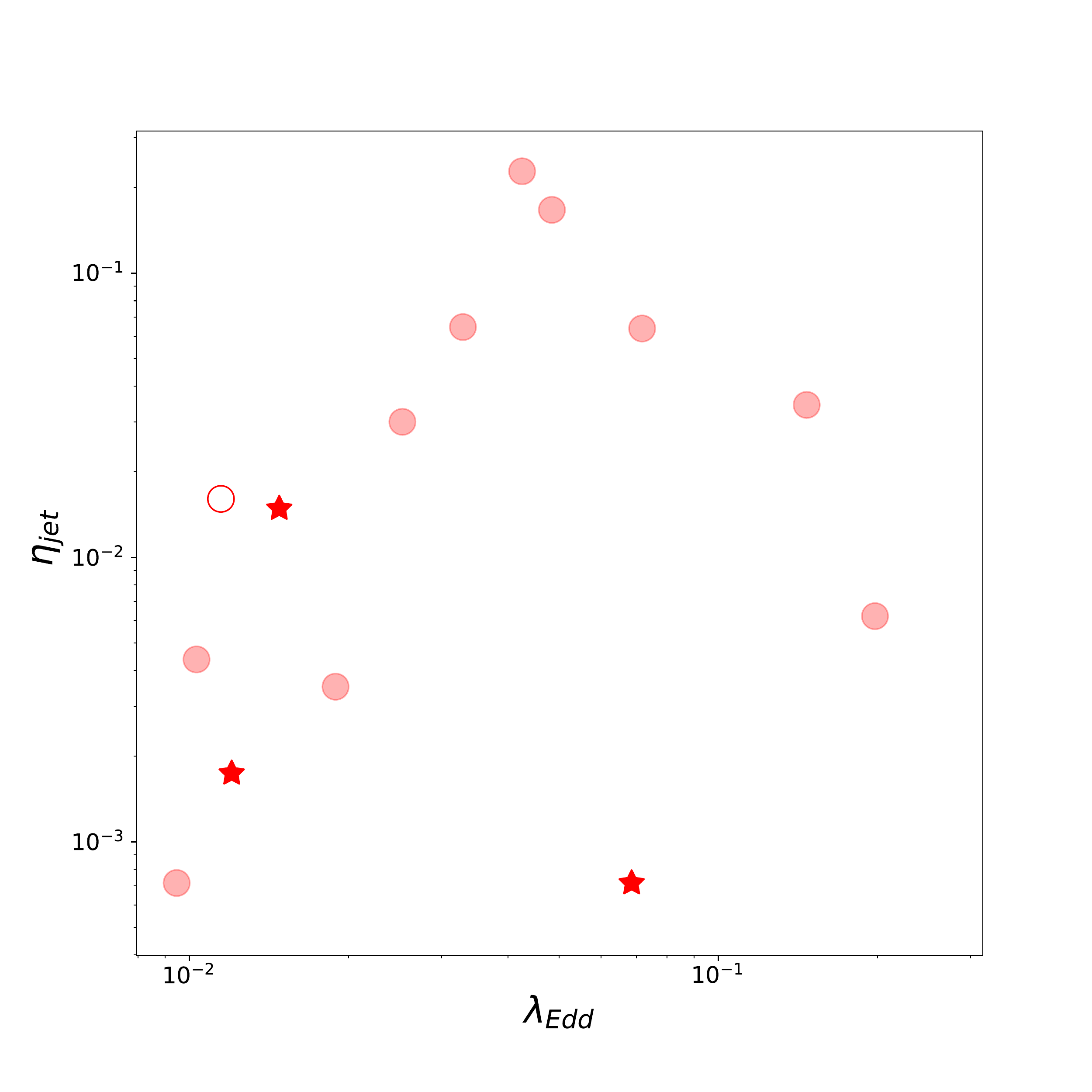}\\
\includegraphics[width=0.8\columnwidth]{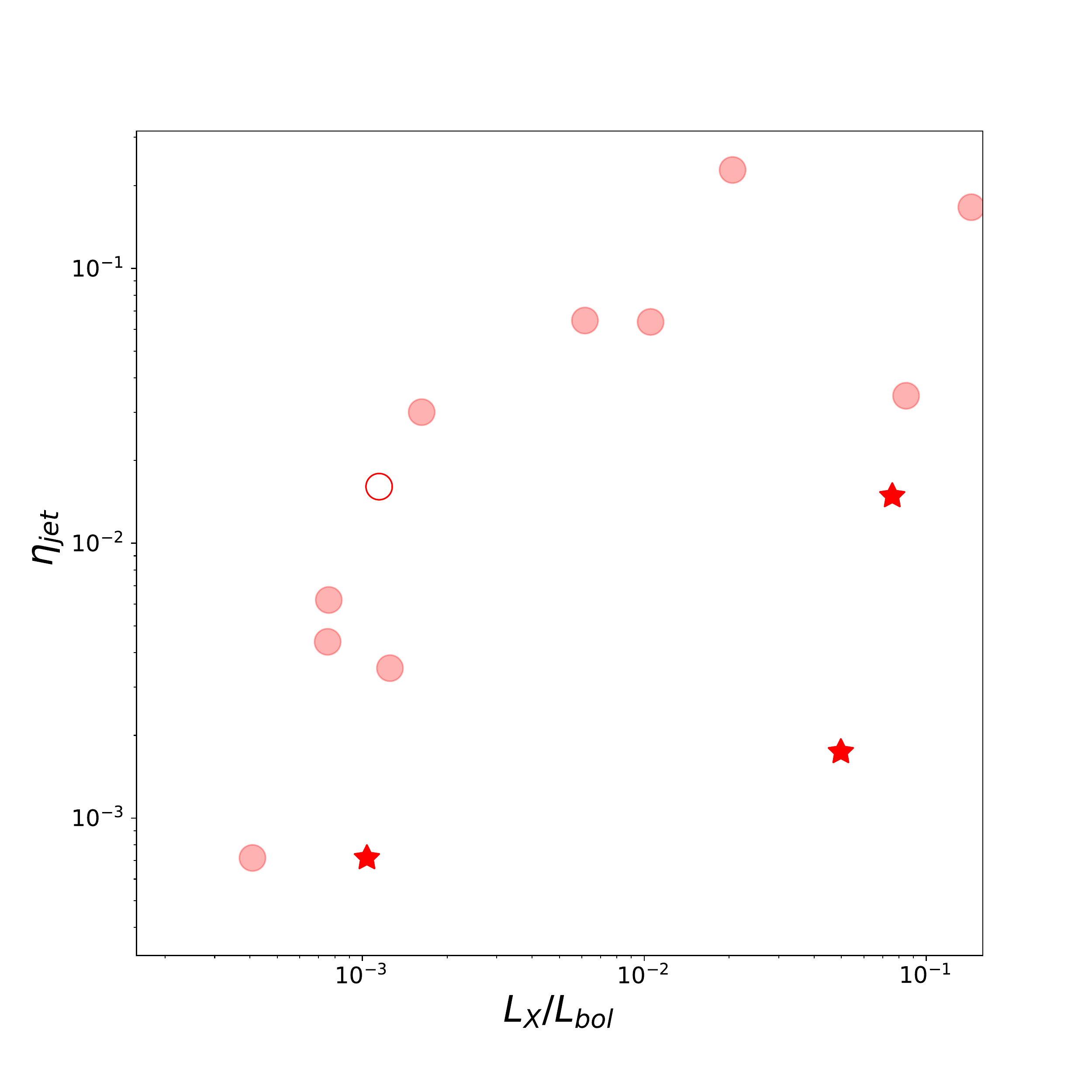}\\
\includegraphics[width=0.8\columnwidth]{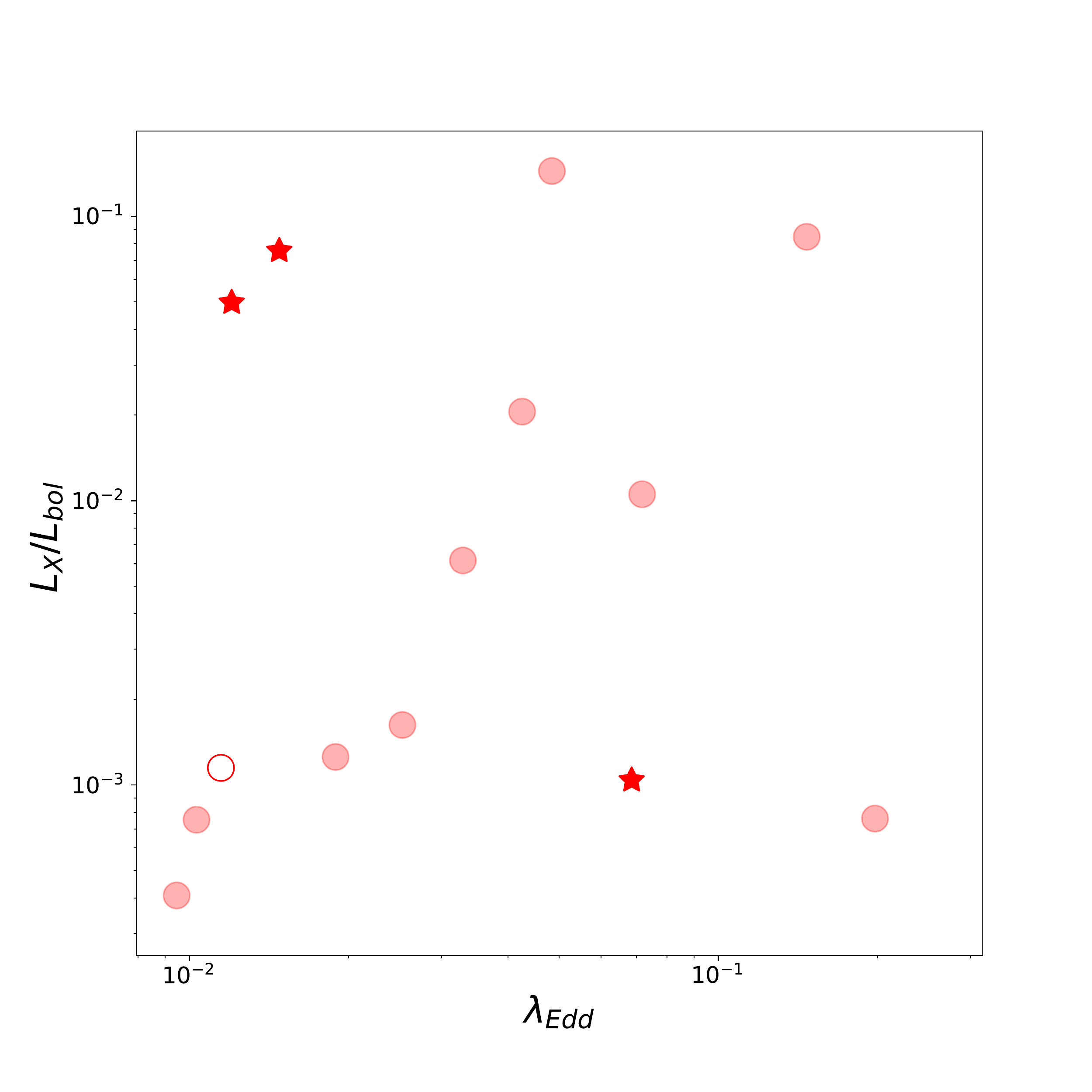}
\caption{Distribution of our CSO sample in the three-dimensional space of the jet production efficiency, $\eta_{\rm jet} \equiv P_{\rm j}/\dot{M}_{\rm acc} c^2$, the accretion rate, $\lambda_{\rm Edd} \equiv L_{\rm bol}/L_{\rm Edd}$, and the X-ray luminosity in units of the bolometric disk luminosity, $L_{\rm X}/L_{\rm bol}$. The empty red circle marks 0116+319, while the Compton-thick sources 2021+614, 1511+0518, and 1404+286, are denoted by red stars. \label{fig:bol}}
\end{figure}

First, we look at the distribution corresponding to the above-mentioned parameters expressed in units of the Eddington luminosity, as presented in Figure \,\ref{fig:Edd}. Here the analyzed sources are represented by filled red circles, except for 0116+319, marked by the empty red circle, for which only the upper limit to the X-ray luminosity could be estimated (see Table \,\ref{table:info}), and the sources confirmed as Compton-thick, denoted in the figure by red stars, namely 2021+614, 1511+0518, and 1404+286 \citep[see][]{Siemiginowska08,Sobolewska19a,Sobolewska19b}. The jet power in the sample spans a relatively wide range from $P_{\rm j}/L_{\rm Edd} \lesssim 10^{-4}$ up to $\sim 10^{-1}$, and the same is true for the X-ray luminosity that varies from $L_{\rm X}/L_{\rm Edd} \lesssim 10^{-5}$ up to $\sim 10^{-2}$. On the contrary, the accretion rate defined in a standard manner as
\begin{equation}
\lambda_{\rm Edd} \equiv \frac{L_{\rm bol}}{L_{\rm Edd}} \equiv \frac{L_{\rm bol}}{4 \pi G \mathcal{M}_{\rm BH} m_{\rm p} c / \sigma_{\rm T}} \, ,
\end{equation}
is distributed within a narrow range, from $\lambda_{\rm Edd} \sim 0.01$ (1718--649) up to $\sim 0.2$ (1934--638). The other feature to notice in the figure, is a correlation between $P_{\rm j}/L_{\rm Edd}$ at lower accretion rates $\lambda_{\rm Edd} \leq 0.05$, followed by a saturation in the jet power for larger values $\lambda_{\rm Edd}> 0.05$, with the only object not complying to the rule being 1404+286, for which however the jet power may be significantly over-estimated (see the previous section). A similar trend can be noted in the $L_{\rm X}/L_{\rm Edd} - P_{\rm j}/L_{\rm Edd}$ plane, except that here all the Compton-thick sources display a different behaviour.

\begin{figure*}
\centering
\includegraphics[width=2.\columnwidth]{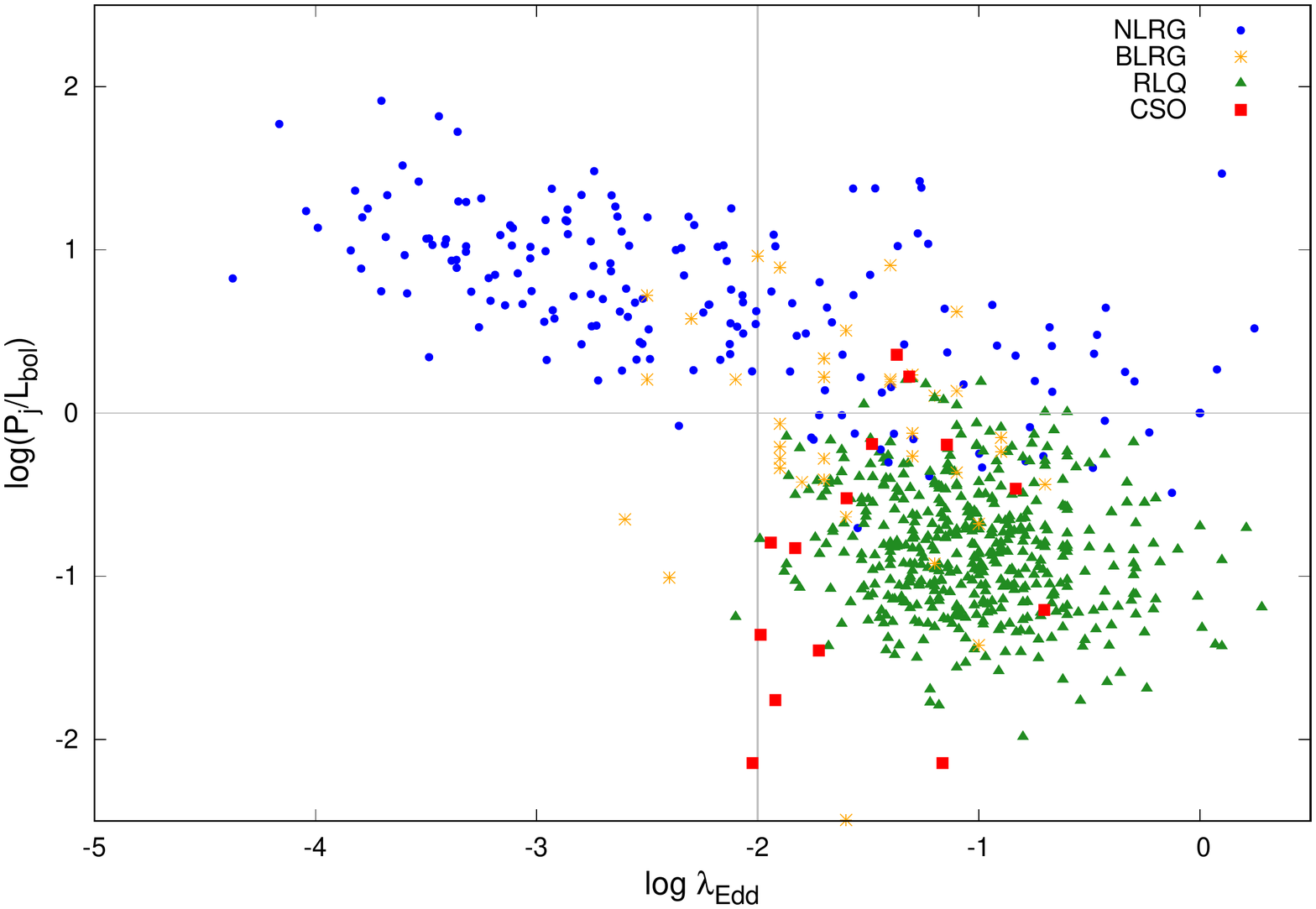}
\caption{Distribution in the $\lambda_{\rm Edd} - P_{\rm j}/L_{\rm bol}$ plane, of the CSOs from our sample (red squares), together with different types of AGN as analyzed by \citet{Rusinek17}, including narrow-line radio galaxies (NLRGs; blue dots), broad-line radio galaxies (BLRGs; orange stars), and radio-loud -- or FR\,II -- quasars (RLQ; green triangles). \label{fig:quasars}}
\end{figure*}

The limited size of the sample precludes any detailed analysis of correlation patterns more complex than linear. Therefore,  we performed the Pearson product-moment correlation coefficients by considering all the sources \emph{except} for the Compton-thick ones, and derived coefficients $r=0.752$ ($p$-value $p=0.008$) for  $P_{\rm j}/L_{\rm Edd} - \lambda_{\rm Edd}$, $r=0.899$ ($p=1.6 \times 10^{-4}$) for $P_{\rm j}/L_{\rm Edd} - L_{\rm X}/L_{\rm Edd}$, and $r=0.779$ ($p=0.005$) for  $L_{\rm X}/L_{\rm Edd} - \lambda_{\rm Edd}$; in all the cases the correlations are positive and statistically significant, according to a confidence level that we fix to $99\%$. This was further supported by the Kendall rank correlation analysis, returning respectively $\tau = 0.6$ ($p=0.0099$), $\tau = 0.855$ ($p=4.6 \times 10^{-5}$), and $\tau = 0.745$ ($p=0.00076$).

Figure \,\ref{fig:bol} shows the distribution of the above source parameters, when the X-ray luminosity is expressed in the units of the bolometric disk luminosity, $L_{\rm X}/L_{\rm bol}$, and the jet kinetic power is expressed in terms of the jet production efficiency, 
\begin{equation}
\eta_{\rm jet} \equiv \frac{P_{\rm j}}{\dot{M}_{\rm acc} c^2} \simeq \frac{P_{\rm j}}{10 \, L_{\rm bol}} \, ,
\end{equation}
where we assumed the standard $10\%$ radiative efficiency of the accretion disks, $L_{\rm bol} \simeq 0.1 \, \dot{M}_{\rm acc} c^2$, as appropriate for the high accretion rates $\lambda_{\rm Edd} \geq 0.01$. This change of the units does not modify the overall picture, except that now the jet production efficiency $\eta_{\rm jet}$ does not really saturate for $\lambda_{\rm Edd}> 0.05$, but decreases with increasing $\lambda_{\rm Edd}$ (again, disregarding 1404+286). When performing a Person correlation analysis on all the data points in the sample excluding the Compton-thick sources, we obtained correlation coefficients $r=0.441$ ($p=0.174$) for  $\eta_{\rm jet} - \lambda_{\rm Edd}$, $r=0.810$ ($p=0.0016$) for $\eta_{\rm jet} - L_{\rm X}/L_{\rm bol}$, and $r=0.535$ ($p=0.089$) for  $L_{\rm X}/L_{\rm bol} - \lambda_{\rm Edd}$. We therefore found a significant, positive correlation only in the case of the $\eta_{\rm jet} - L_{\rm X}/L_{\rm bol}$ dependence; this was again supported by the Kendall rank correlation analysis, returning $\tau = 0.382$ ($p=0.121$), $\tau = 0.709$ ($p=0.0016$), and $\tau = 0.6$ ($p=0.01$), respectively.

We emphasize that the jet production efficiency for the analyzed CSOs spans a range from $\eta_{\rm jet} \lesssim 10^{-3}$ (1718--649), up to $\sim 0.2$ (0108+388) at most. This means that the newly-born radio galaxies in our sample do not reach the maximum level of jet production efficiency, as could be expected in the case of  magnetically-arrested geometrically-thick disks around maximally spinning black holes \citep[$\eta_{\rm jet} \gtrsim 1$; see][]{Tchekhovskoy11}, unless the jet kinetic powers are significantly under-estimated here (by orders of magnitude!). We note, on the  other hand, that the jet production may proceed with a much lower efficiency ($\eta_{\rm jet} < 1$) in the case of geometrically-thin disks, even in the regime of a strong magnetization \citep[see][]{Avara16}.

\begin{figure}
\centering
\includegraphics[width=1.065\columnwidth]{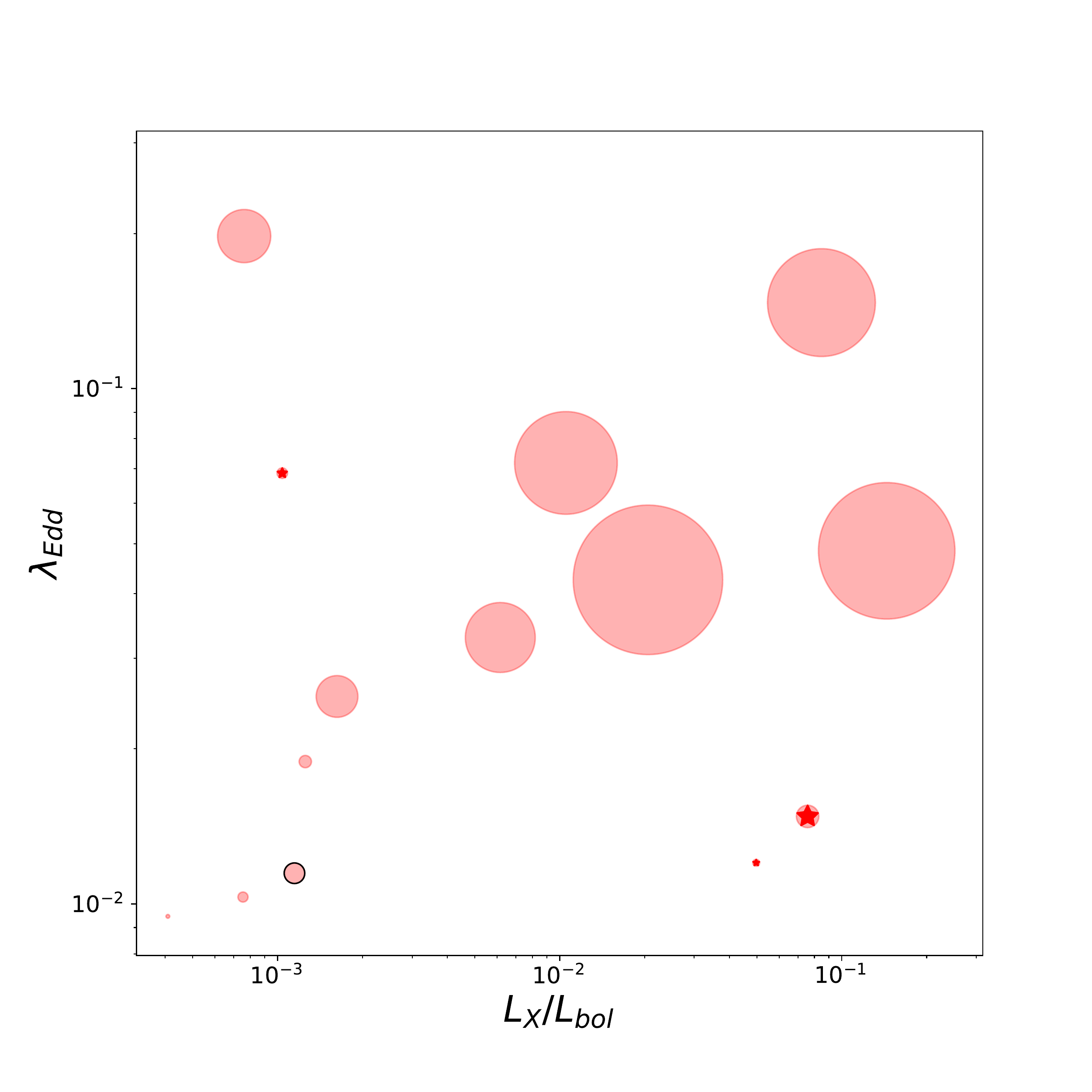}
\includegraphics[width=1.065\columnwidth]{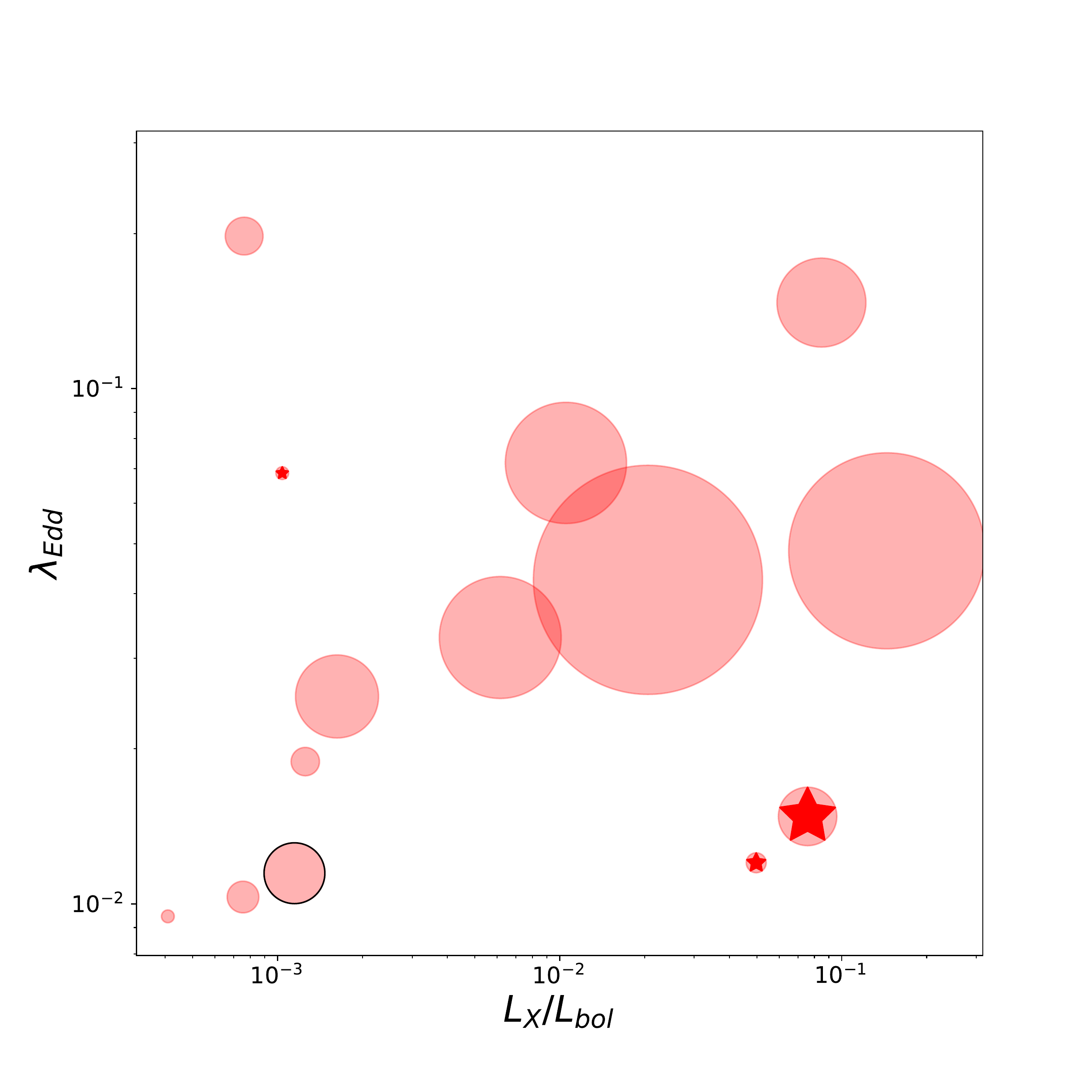}
\caption{Dependence of the jet production efficiency on the combination of the accretion-related parameters $\lambda_{\rm Edd}$ and $L_{\rm X}/L_{\rm bol}$, for the CSOs from the analyzed sample. The jet power is coded by the size of the symbols (so that larger jet kinetic luminosities are denoted by larger circles), and expressed in Eddington units $P_{\rm j}/L_{\rm Edd}$ (left panel), or in units of mass accretion rate $\eta_{\rm jet}$ (right panel). Compton-thick sources are marked by red stars. \label{fig:states}}
\end{figure}

Quite surprisingly, in the $\lambda_{\rm Edd} - \eta_{\rm jet}$ plane, the CSOs from our sample occupy the same locus as ``double-lobed'' (i.e., Fanaroff-Riley type II) radio quasars, discussed in a similar context by, e.g., \citet{Velzen13}, and more recently also \citet{Rusinek17}. In order to illustrate this finding, in Figure \,\ref{fig:quasars}, we plot all the sources from the \citeauthor{Rusinek17} sample --- including narrow-line radio galaxies, broad-line radio galaxies, and radio quasars --- along with our targets (denoted in the figure by red squares; note that here on the $y$-axis we plot simply $P_{\rm j}/L_{\rm bol}$ instead of $\eta_{\rm jet}$). The jet kinetic power estimates in \citeauthor{Rusinek17} are based on scaling given in equation\,\ref{eq:Will} above, while for our targets we are considering the minimum jet power estimates given in equation\,\ref{eq:min}. If, instead of such lower limit values, one uses equation\,\ref{eq:Will}, the distribution of our CSOs would overlap quite closely with that of BLRGs from the \citeauthor{Rusinek17} sample.

One should point out, however, that the majority of the sources analyzed by \citet{Rusinek17} posses extended radio structures, with estimated ages of several Myrs, and so the jet kinetic powers derived for these sources by means of a time-averaging calorimetric relation \citep[as in][equation\,\ref{eq:Will} above]{Willott99} may not correspond to the accretion rates estimated from the currently available optical spectroscopic data. On the other hand, in the cases of the youngest radio galaxies, as considered here (ages $<3$\,kyr), the currently observed radio luminosities of compact jets and lobes \emph{and} the currently observed optical luminosities of the accretion disks, likely correspond to the same episode of the AGN activity.

In Figure \,\ref{fig:states} we visualize the dependence of the jet production efficiency on the combination of the accretion-related parameters $\lambda_{\rm Edd}$ and $L_{\rm X}/L_{\rm bol}$. In both panels of the figure, the $x$-axes give the $L_{\rm X}/L_{\rm bol}$ ratio: under the assumption that the observed X-ray luminosities are good proxies for the disk coronal emission, this ratio corresponds to different states of the accretion disks, namely ``hard'' states at large $L_{\rm X}/L_{\rm bol}$ values, and ``soft'' states otherwise. The $y$-axes in the figure denote the accretion rate $\lambda_{\rm Edd}$, i.e. the disk luminosity expressed in Eddington units. The jet power in the figure is coded by the size of the symbols, so that larger jet kinetic luminosities are denoted by larger circles. In the left panel of the figure, the jet power is expressed in the Eddington units, $P_{\rm j}/L_{\rm Edd}$, and in the right panel in the units of the mass accretion rate, $\eta_{\rm jet}$. As before, Compton-thick sources are marked by red stars.

Galactic X-ray Binaries (XRBs) are known to trace a ``q-shaped hysteresis'' (in the counter-clockwise direction) on exactly this kind of ``hardness--intensity'' diagram for the accretion states\footnote{Note that accretion disks in XRBs are much hotter than accretion disks in AGN, due to the inverse scaling between disk temperatures and black hole masses. As a result, for XRBs the disk continua are pronounced in soft X-rays, and so the coronal-to-disk luminosity ratio can be expressed convininetly as the X-ray hardness ratio, while the accretion rate is simply proportional to the soft X-ray intensity.} \citep{Homan05,McClintock06}: starting at the low-accretion-rate but hard state (bottom-right corner of the panels), the sources increase their disk intensity, moving vertically up in the diagram up to the point when the accretion rate saturates at the maximum value; from that point, the sources evolve horizontally to the left, i.e. toward the soft state, until the minimum hardness level is reached (top-left corner of the panel), they then decrease their accretion rate, and finally returning to the starting point. This spectral evolution is accompanied by changes in the jet production efficiency \citep{Fender04}: one typically sees steady low-power jets with luminosities scaling with the disk intensities during the source evolution in the hard states, followed by an episode of high-power but highly intermittent jets during the transition to the high/soft state, and finally a suppression of the jet production during the low/soft states. 

The SMBHs masses in AGN are orders of magnitude ($>10^6$ times) larger than the black hole masses in XRBs. Thus, all the related accretion-related timescales are correspondingly longer, implying that one cannot witness in real time any track traced by a given AGN on the accretion hardness--intensity diagram. However, in a well defined sample of AGN, the distribution of the sources in a diagram may reveal an analogous evolutionary trend in a statistical manner \citep[see, e.g.,][]{Sobolewska11}. Even though the distribution of our young radio galaxies in Figure \,\ref{fig:states} does not reveal a clearly defined ``q-shaped hysteresis'', it shows an interesting dependence of the jet production efficiency on the combination of the coronal and disk intensities, with the jets being produced most efficiently during the high/hard states, and suppressed during the soft states.

\section{Conclusions}
\label{sec:conclusions}

In this paper, we studied the jet production efficiency in the sample of the 17 confirmed young radio galaxies of the CSO type for which the redshifts are known, the kinematic ages have been estimated directly from the monitoring radio data, and the nuclear X-ray fluxes have been measured from high-angular resolution observations with either {\it Chandra} or XMM-{\it Newton}. The analyzed sample is unique: (i) our targets are the most robust and unambiguous examples of the youngest radio galaxies, with linear sizes $\lesssim 300$\,pc and corresponding ages $\lesssim 3,000$\,yr; (ii) their spectra are free of relativistic beaming effects; (iii) the currently observed radio and optical luminosities of their compact lobes and accretion disks, respectively, correspond to the same episode of the AGN activity.

For these targets, we have analyzed the available SDSS spectra, and estimated the bolometric luminosities of the accretion disks, $L_{\rm bol}$, using the narrow $H\beta$ line, and the black hole masses, $\mathcal{M}_{\rm BH}$, from the measured stellar velocity dispersion. When SDSS spectra were not available, we have reviewed the other optical data presented in the literature, gathering the estimates of $L_{\rm bol}$ based on either the narrow $H\beta$ line or the [OIII] line, and the estimates of $\mathcal{M}_{\rm BH}$ based on either the stellar velocity dispersion, the galaxy bulge luministy, or the absolute R-band optical magnitude of the host. 

We have estimated the jet kinetic power for the objects from our list, $P_{\rm j}$, by means of the calorimetric scaling relation with the observed $1.4$\,GHz radio fluxes, established for the evolved radio sources by \citet{Willott99}. We have also derived the minimum jet powers, utilizing the robust kinematic age estimates, and assuming energy equipartition between magnetic field and radio-emitting electrons within compact lobes. In the majority of the cases, the minimum jet kinetic luminosities turned out to be lower by one order of magnitude than the corresponding jet powers derived from the radio scaling relation. We concluded that the latter values over-estimate the jet power in young radio sources, due to the enhanced radiative efficiency of compact radio-emitting jets and lobes, as discussed by \citet{Tadhunter11,Dicken12}.

The estimated minimum jet power, along with the known kinematic age of the analyzed sources, provide the lower limit to the total energy deposited by the jets within the central regions ($<1$\,kpc) of the host galaxy. This energy varies from $P_{\rm j} \times \tau_{\rm j} \sim 10^{52}$\,erg in the case of 1718--649, up to $\sim 10^{56}$\,erg in the case of 1607+26. These values imply that at least the most powerful young radio sources may provide a significant impact on the evolution of galaxy bulges of their hosts \citep[see in this context][and references therein]{Tadhunter16}.

We have also investigated the distribution of our CSO sample in the three-dimensional space defined by the bolometric luminosity $L_{\rm bol}$, the nuclear X-ray luminosity $L_{\rm X}$ (considered here as a proxy for the radiative output of the accretion disk coronae), and the minimum jet kinetic luminosity $P_{\rm j}$, expressing the former parameter in terms of the accretion rate $\lambda_{\rm Edd} \equiv L_{\rm bol}/L_{\rm Edd}$, and the latter two parameters either in Eddington units, or in units of the bolometric disk luminosity. The main findings from this analysis can be summarized as follows:
\begin{itemize}
\item[i)] the accretion rate $\lambda_{\rm Edd}$ in our sample is distributed within a narrow range from $\lambda_{\rm Edd} \sim 0.01$ up to $\sim 0.2$ (unlike the normalized jet power or the normalized X-ray luminosity, which both span a much wider range);
\item[ii)] the normalized jet power $P_{\rm j}/L_{\rm Edd}$ correlates with the accretion rate at $\lambda_{\rm Edd} \leq 0.05$, and saturates (or even decreases) at larger values $\lambda_{\rm Edd}> 0.05$;
\item[iii)] the jet production efficiency $\eta_{\rm jet} \equiv P_{\rm j}/\dot{M}_{\rm acc} c^2$ spans a range from $\eta_{\rm jet} \lesssim 10^{-3}$ up to $\sim 0.2$ at most, meaning that the newly-born radio galaxies in our sample do not reach the highest possible level of the jet production efficiency, as could be expected in the case of magnetically arrested disks around maximally spinning black holes;
\item[iv)] we see an interesting diversification in the jet production efficiency on the hardness--intensity diagram $L_{\rm X}/L_{\rm bol} - \lambda_{\rm Edd}$ characterizing the accretion state of young radio galaxies, somewhat analogous to that observed in Galactic XRBs, with the jets being produced most efficiently during the high/hard states, and suppressed during the soft states.
\end{itemize}

\acknowledgments
This work (AW, \L S, and EK) was supported by the Polish NCN grant 2016/22/E/ST9/00061. 
L.O. acknowledges partial support from the INFN Grant InDark and the grant of the Italian Ministry of Education, University and Research (MIUR) (L. 232/2016) ``ECCELLENZA1822 D206 - Dipartimento di Eccellenza 2018-2022 Fisica'' awarded to the Dept. of Physics of the University of Torino.
The authors acknowledge the anonymous referee for her/his comments and suggestions.

\begin{turnpage}
\begin{deluxetable*}{lllllllllllllll}
\tabletypesize{\scriptsize}
\tablecaption{Dataset with the values used and/or estimated in this work for the sample of 17 CSOs with measured redshifts, kinematic ages, and with high resolution X-ray observations. \label{table:info}}
\tablehead{
\colhead{source} & \colhead{$z$} & \colhead{$LS$} & \colhead{$\tau$} & \colhead{$d_{\rm L}$} &  \colhead{$L_{2-10\,\rm{keV}}$}  & \colhead{$\log \frac{\mathcal{M}_{\rm BH}}{M_{\odot}}$} & \colhead{method} &  \colhead{$L_{\rm bol}$} &  \colhead{method} &  \colhead{$\log \lambda_{\rm Edd}$} &  \colhead{$L_{1.4\,\rm{GHz}}$} &  \colhead{$L_{5\,\rm{GHz}}$} &  \colhead{$P_{\rm j}$} &  \colhead{$\log \eta_{\rm jet}$} \\
\colhead{} & \colhead{} & \colhead{[pc]} & \colhead{yrs} & \colhead{[Mpc]} & \colhead{[$10^{42}$\,erg/s]} & \colhead{} & \colhead{} & \colhead{[$10^{44}$\,erg/s]} & \colhead{} &  \colhead{} & \colhead{[$10^{24}$\,W/Hz]} & \colhead{[$10^{42}$\,erg/s]} & \colhead{[$10^{43}$\,erg/s]} & \colhead{}\\
\colhead{(1)} & \colhead{(2)} & \colhead{(3)} & \colhead{(4)} & \colhead{(5)} & \colhead{(6)} & \colhead{(7)} & \colhead{(8)} & \colhead{(9)} & \colhead{(10)} & \colhead{(11)} & \colhead{(12)} & \colhead{(13)} & \colhead{(14)} & \colhead{(15)}
}
\tiny
\startdata 
0035+227 & 0.096 & 21.8 & 450 $^a$ & 418 & 0.75 [S16] & 8.4 &  $\sigma_\star$ [S12] & 5.98 & $H\beta$ [S12]  &  --1.72 & 12.89 $^b$ & 0.26 $^e$ & 2.1 & --2.46\\
0108+388 & 0.669 & 22.7 & 404 $^a$ & 3907 & 70 [T09] & 7.9 &  $M_{\rm R}$ [W09] & 4.85 & [OIII] [W09] &  --1.31 & 783.5 $^b$ & 115 $^c$ & 81 & --0.78\\
0116+319 & 0.059 & 70.1 & 501 $^a$ & 255 & $\rm{<1.0} ^{\dagger\dagger}$ [S16] & 8.8 &  $L_{\rm blg}$ [W10] & 8.71 & [OIII] [W09]  &  --1.94 & 20.51 $^b$ & 0.62 $^d$ & 14 & --1.79\\
0710+439 & 0.518 & 87.7 & 932 $^b$ & 2868 & 394 [S16] & 8.4 &  $M_{\rm R}$ [W09] & 46.5 & $H\beta$ [L96]  &  --0.83 & 1980 $^b$ & 82.7 $^c$ & 160 & --1.45\\
1031+567 & 0.460 & 109.0 & 620 $^c$ & 2480 & 22 [T09] & 8.3 &  $ \sigma_\star$ & 10.7 & $H\beta$  &  --1.37 & 1439 $^b$ & 47.1 $^c$ & 244 & --0.64\\
1245+676 & 0.107 & 9.6 & 188 $^a$ & 478 & 0.31$^\star$ [Wa09] & 8.5 &  $\sigma_\star$ [S12] & 4.11 & $H\beta$ [S12] &  --1.99 & 7.878 $^b$ & 0.26 $^e$ & 1.8 & --2.37\\
1323+321 & 0.368 & 278.1 & 1030 $^d$ & 2010 & 59 [T09] & 9.3 & $ \sigma_\star$ [S12] & 62.5 & $H\beta$ [S12] &  --1.50 & 493 $^b$ & 59 $^f$ & 404 & --1.05\\
1404+286 & 0.077 & 7.0 & 219 $^e$ & 336 & 4.5$^\dagger$ [S19b] & 8.6 &  $ \sigma_\star$ & 43.3 & 12\,$\mu$m [K19] &  --1.61 & 11.92 $^b$ & 1.82 $^c$  & 3.1 & --3.15\\
1511+0518 & 0.084 & 7.3 & 300 $^f$ & 370 & 30$^\dagger$ [S16] & 8.6 &  $\sigma_\star$  & 6.03 &  SED [T13] &  --1.92 & 1.075 $^b$ & 0.442 $^c$  & 1.05 & --2.76\\
1607+26 & 0.473 & 240 & 2200 $^g$ & 2569 & 37.9 [S16] & 8.6 &  $\sigma_\star$ & 36.0 & $H\beta$  &  --1.14 & 4213 $^b$ & 68.3 $^c$ & 230 & --1.20\\
1718--649 & 0.014 & 2.0 & 91 $^a$ & 60 & 0.15 [S16] & 8.5  &  $L_{\rm blg}$ [W10] & 3.77 & $H\beta$ [P96]  &  --2.02 & 1.545 $^a$ & 0.0943 $^c$  & 0.27 & --3.14\\
1843+356 & 0.763 & 22.3 & 180 $^b$ & 4612 & 56 [S16] &  --- &  --- & --- & ---  &  --- & 2183 $^b$ & 104 $^c$  & 170 & ---\\
1934--638 & 0.183 & 85.1 & 1603 $^a$ & 845 & 6 [S19a] & 8.5 &  $M_{\rm R}$ [W09] & 78.8 & $H\beta$ [R16]  &  --0.70 & 1280 $^a$ & 27.8 $^c$ & 49 & --2.20\\
1943+546 & 0.263 & 107.1 & 1308 $^a$ & 1285 & 7.31 [S16] & 8.5 &  $\sigma_\star$ [S12] & --- & ---  &  --- & 346.5 $^b$ & 9.29 $^e$ & 43 & ---\\
1946+708 & 0.101 & 39.4 & 1261 $^a$ & 444 & 12 [S19a] & 8.5 &  $L_{\rm blg}$ [W10] & --- &  ---  &  --- & 22.49 $^b$ & 0.76 $^c$  & 3 & ---\\
2021+614 & 0.227 & 16.1 & 368 $^b$ & 1086 & 112$^\dagger$ [S19a] & 8.9  &  $M_{\rm R}$ [W09] & 14.8 & [OIII] [W09] &  --1.83 & 301.1 $^b$ & 21.2 $^d$  & 22 & --1.84\\
2352+495 & 0.238 & 117.3 & 3003 $^b$ & 1143 & 13 [T09] & 8.4 &  $M_{\rm R}$ [W09] & 8.0 & $H\beta$ [L96]  &  --1.60 & 360.5 $^b$ & 11.7 $^e$  & 24 & --1.52
\enddata
\tablecomments{\\
{Column (1): source name;}\\ 
{Column (2): redshift;} \\
{Column (3): linear size; see reference in column 4;} \\
{Column (4): kinematic age; references: $^a$ \citet{Giroletti09}, $^b$ \citet{Polatidis03}, $^c$ \citet{Taylor00}, $^d$ \citet{An12a}, $^e$ \citet{Luo07}, $^f$ \citet{An12b}, $^g$ \citet{Nagai06};} \\
{Column (5): luminosity distance;}\\
{Column (6): absorption-corrected X-ray luminosity; references: [S16] \citet{Siemiginowska16}, [T09] \citet{Tengstrand09}, [Wa09] \citet{Watson09}, [S19a] \citet{Sobolewska19a}, [S19b] \citet{Sobolewska19b}; $^{\dagger}$ value corresponding to the Compton-thick scenario; $^{\dagger\dagger}$ estimated from the 0.5-2.0\,keV upper limit assuming photon index $\Gamma$ = 2.0; $^{\star}$ calculated based on the 4.5--12.0\,keV PN flux;}\\
{Column (7): SMBH mass; see references in column 6;}\\
{Column (8): method for the SMBH mass estimates; references: [W10] \citet{Willett10}, [W09] \citet{Wu09}, [S12] \citet{Son12}, otherwise this work;}\\
{Column (9): accretion disk bolometric luminosity; see references in column 8;}\\
{Column (10): method used to derive the bolometric disk luminosity; references to the spectroscopic data: [P96] \citet{Polletta96}, [L96] \citet{Lawrence96}, [R16] \citet{Roche16}, [S12] \citet{Son12}, [W09] \citet{Wu09}, [T13] \citet{Trichas13}, [K19] \citet{Kosmaczewski19}, otherwise this work;}\\
{Column (11): accretion rate $\lambda_{\rm Edd} \equiv L_{\rm bol}/L_{\rm Edd}$;}\\
{Column (12): 1.4\,GHz luminosity spectral density; references: $^a$ \citet{Tingay03}, $^b$ \citet{Condon98};}\\
{Column (13): 5\,GHz monochromatic luminosity; references: $^c$ \citet{Labiano07}, $^d$ \citet{Horiuchi04}, $^e$ \citet{Gregory96};}, $^f$ \citet{Stanghellini98}\\
{Column (14): minimum jet kinetic power;}\\
{Column (15):  jet production efficiency $\eta_{\rm jet} \equiv P_{\rm j}/\dot{M}_{\rm acc} c^2$.}
}
\end{deluxetable*}
\end{turnpage}
\end{document}